\documentclass[8pt,a4paper]{article}
\usepackage[utf8]{inputenc}
\usepackage{amsmath}
\usepackage{amsfonts}
\usepackage{amssymb}
\usepackage{graphicx}
\usepackage{xcolor}
\usepackage{url}
\usepackage{braket}
\usepackage{tikz}
\usepackage[left=3cm, right=3cm, top=2cm, bottom=2cm]{geometry}
\usepackage{rotating}
\usepackage{lscape}
\usepackage{hyperref}
\usepackage{lscape}
\usepackage{wrapfig}
\usepackage{float}

\DeclareMathOperator{\sgn}{sgn}
\DeclareMathOperator{\tr}{tr}
\newcommand{\rl}[1]{\left(#1\right)}

\allowdisplaybreaks
\graphicspath{{figures/}}
\allowdisplaybreaks[1]
\begin{document}

	\baselineskip 24pt

\begin{center}

        {\large \textbf{QUANTUM ENTANGLEMENT IN THE ONE-DIMENSIONAL ANYONIC
HUBBARD MODEL}\par}

\end{center}

\vskip .5cm
\medskip

\vspace*{4.0ex}

\baselineskip=18pt

\centerline{\large\rm  Ramadas N, V V Sreedhar}

\vspace*{4.0ex}

\centerline{\it Chennai Mathematical Institute, PLOT H1,  SIPCOT IT Park, 
Siruseri, Chennai, 603103 India} 

\vspace*{1.0ex}

\vspace*{1.0ex}
\centerline{ \href{mailto:ramadasn@cmi.ac.in}{\texttt{ramadasn@cmi.ac.in}}, \href{mailto:sreedhar@cmi.ac.in}{\texttt{sreedhar@cmi.ac.in}} }

\vspace*{5.0ex}

\begin{abstract}

Issues related to quantum entanglement in systems of indistinguishable
particles, as discussed in the information theoretic approach, are extended
to anyonic statistics. Local and non-local measurements discussed in this
framework are carefully analysed in the two-site anyonic Hubbard model which
provides a concrete case-study. The von Neumann entropy, the single-particle 
density matrix, the pair correlation function, and the pseudo-momentum 
distribution function are worked out paying special attention to the 
dependence on the statistics parameter.

\end{abstract}

\section{Introduction}

Interest in the study of anyons -- particles in low spatial dimensions with
more general quantum statistics than bosons and fermions -- has been revived
in the recent times because of their purported role in topological quantum
computation  \cite{kitaev2006anyons} \cite{preskilltqc} \cite{sarma2006topological}.

In a recent paper,  the authors presented a way to characterise the entanglement
between anyons by studying the dependence of the von Neumann entropy on the
parameter characterising anyonic statistics in one dimension \cite{maniQuantumEntanglementOnedimensional2020}.
The model considered, first introduced by Leinaas and Myrheim \cite{leinaasTheoryIdenticalParticles1977},
deals with the quantum mechanics of two indistinguishable point particles on
the real line. Anyonic behaviour follows from identifying the correct
configuration space, and employing Robin boundary conditions at the singular
points. The parameter in the definition of the Robin boundary conditions, which
interpolates between bosons and fermions, plays the role of the statistics
parameter.

The results were obtained by suitably generalising, to the case of anyons, 
the information-theoretic approach to quantum entanglement of bosons and 
fermions, pioneered by Compagno and Lo Franco \cite{franco2016quantum}, who 
applied it to the study of quantum entanglement between two qubits in a 
one-dimensional asymmetric double-well potential.

The second quantization picture of the information-theoretic ideas, also
outlined by Castellini, Compagno and Lo Franco \cite{compagno2018dealing}, is especially useful 
in going beyond bosons and fermions. The results of \cite{maniQuantumEntanglementOnedimensional2020} follow 
as a direct consequence of the generalised commutation relations satisfied 
by the anyon field operators in the Leinaas-Myrheim model \cite{posske2017second}. 
When the statistics parameter tends to zero or infinity, the limiting cases 
corresponding to bosons and fermions, it was found that the two-particle 
states of the Leinaas-Myrheim model had exactly the same entanglement 
properties as the qubits in the asymmetric double well.

A more direct study of the model of qubits in the asymmetric double-well is 
desirable because it would shed light not only on the entanglement between 
the spin degrees of freedom, but also between the spatial modes, and hence 
on anyonic statistics resulting from the exchange of positions. This would 
require a thorough analysis of the solutions of the coupled Schr\"odinger 
equations (more precisely, Pauli equations) for two spin-half particles
in the asymmetric double-well potential, and a suitable application of
boundary conditions. Needless to say, this is a cumbersome exercise.

The essential physics of this system is easily captured in a much simpler
model {\it viz.}, the two-site Hubbard model \cite{hubbardElectronCorrelationsNarrow1963}  . This model is
defined by a Hamiltonian which describes the dynamics of electrons on a
one-dimensional lattice. In the simplified two-site model, the left and
right lattice sites correspond to the left and right minima of the double-well
potential. The term in the Hamiltonian representing the hopping of an electron
from one site to the nearest neighbour site corresponds to tunneling
between the two minima. The on-site potential terms in the Hamiltonian
correspond to the depth of the minima in the double-well, and the spin
degrees of freedom of the electrons are the qubits. By switching between
commutation and anti-commutation relations for the creation and annihilation
operators, one can easily go from bosons to fermions. Moreover, generalisation
to anyonic statistics is a straightforward exercise comprising of using the
appropriate generalised commutation relations for the creation and
annihilation operators. Because it is two-site model, it is amenable to
exact diagonalisation. We can also couple the particles to external fields.
As a by-product of the exact diagonalisation, we can compute the statistics
parameter dependence of the one-particle density matrix, the pair correlation
function, and the quasi-momentum distribution of anyons. Although we will not
be doing so in this paper, it is also straightforward to generalise the methods
in this paper to allow for an arbitrary number of particles hopping on a 
lattice of an arbitrary number of sites, as also for qudits and other 
higher-spin particles.

The remainder of this paper is organised as follows. In Sec. \ref{inftate}
we briefly review the information theoretic approach to entanglement of
indistinguishable particles developed by Compagno and Lo Franco \cite{franco2016quantum}, discuss the notions of local and nonlocal 
measurements,
and offer some additional insight into their ideas by connecting them to more
familiar notions of singular value decomposition used in the Schmidt
decomposition \cite{nielsenQuantumComputationQuantum2002} for distinguishable particles.

In Sec. \ref{theahm}, we introduce an anyonic generalisation of the usual
Hubbard model in one dimension and solve the two-site case, by exact 
diagonalisation, for its eigenvalues and eigenvectors.
The model is also discussed in the presence of an external field.

In Sec. \ref{ente}, we apply the information theoretic approach to
entanglement of indistinguishable particles to the states of the two-site
anyonic Hubbard model. In particular, it is shown that the von Neumann
entropy of anyons varies with time, and the statistics parameter, in the 
presence of an external field. We also discuss entanglement extraction, introduced, and applied to
the qubit states in the asymmetric double-well potential,  by Compagno 
and Lo Franco in \cite{franco2016quantum}, and examine the 
corresponding results in the two-site anyonic Hubbard model.

In Sec. \ref{corf} we present
 a calculation of the one-particle density
matrix and the pair correlation function in the two-site anyonic Hubbard 
model with special emphasis on their dependence on the statistics parameter.

In Sec. \ref{quamd}, we present results for the quasi-momentum distribution
functions of anyons as a function of the statistics parameter.

We conclude by presenting a summary and outlook of the results, and a
set of appendices containing the calculational details.

\section{ Information Theoretic Approach to Quantum Entanglement} \label{inftate}

In quantum mechanics, various measures are used to quantify entanglement
between subsystems of a composite system. For distinguishable particles,
there are well-defined entanglement measures. For indistinguishable particles, 
however, one needs to exercise care in defining these measures. Many 
approaches \cite{franco2016quantum} \cite{balachandran2013algebraic} \cite{ghirardi2005identical} \cite{schliemann2001quantum} \cite{pavskauskas2001quantum}
\cite{wiseman2003entanglement}\cite{killoran2014extracting}\cite{franco2018indistinguishability}\cite{benatti2014entanglement} have been developed to deal with this issue, of which 
we prefer to use the method of Compagno and Lo Franco based on ideas from 
information theory \cite{franco2016quantum}. We briefly review the information theoretic approach to indistinguishable particles here. Some interesting experimental results pertaining to this formalism can be found in \cite{sunExperimentalQuantumEntanglement2020} \cite{sunExperimentalQuantumPhase2021}.

In the usual way, indistinguishable particles are labeled first as if they
were distinguishable, and then their wavefunctions are symmetrized or 
anti-symmetrized with respect to the labels to get the wavefunctions of 
bosons or fermions. These labels cause confusion while defining entanglement 
of indistinguishable particles. Therefore, an approach which does not label 
indistinguishable particles is desirable. In the information-theoretic 
approach, the state of indistinguishable particles is treated holistically,
as a single entity.

It is useful to illustrate the above points in the simple case of a 
two-particle state $\ket{\phi,\psi}$.  It is important to stress here that 
$\phi,\psi$ do not refer to particle labels. The fact that there are two 
arguments in the ket suggests that we are dealing with a two-particle state. 

Assume a physical process where the two-particle state $\ket{\phi,\psi}$
is transformed to the state $\ket{\varphi,\zeta}$. Since the particles are
indistinguishable, the two-particle probability amplitude for the process 
is given by the inner product
\begin{align}
\begin{aligned}
\braket{\varphi,\zeta| \phi,\psi} := \braket{\varphi|\phi} \braket{\zeta|\psi} 
+\eta \braket{\varphi|\psi} \braket{\zeta|\phi}
\end{aligned}
\end{align}
where $ \eta=\pm 1 $, corresponding to bosons or fermions.

A connection with single-particle states $\ket{\phi}$ and $\ket{\psi}$ 
requires the introduction of the inner product between between two-particle 
and one-particle states which, in the present case, is defined by 
\begin{align}
\begin{aligned}
\braket{\psi|\cdot |\varphi_1,\varphi_2} := \braket{\psi| \varphi_1,\varphi_2} 
= \braket{\psi|\varphi_1} \ket{\varphi_2}+\eta \braket{\psi|\varphi_2}
\ket{\varphi_1}
\end{aligned}
\end{align}
The above equation gives a one-particle state by projecting the two-particle
state $\ket{\varphi_1,\varphi_2}$ on to the one-particle state $\ket{\psi}$.
This inner product between states with different dimensionality can be used to
define partial trace and thus the reduced density matrix.

Consider now a general, normalized $N-$particle state $ \ket{\Phi}$. One can 
define the inner product between $ \ket{\Phi}$ and a single-particle state 
$\ket{\psi}$ as follows: Consider a basis $\lbrace\ket{k} \rbrace $ of the 
single- particle Hilbert space. The normalized $(N-1)$-particle state obtained 
by projecting $\ket{\Phi} $ onto $ \ket{k} $ is given by
\begin{align}
\begin{aligned}
\ket{\phi_k} = \frac{\braket{k|\Phi}}{\sqrt{\braket{\Phi|\Pi^{(1)}_k|\Phi}}}
\end{aligned}
\end{align}
where $ \Pi^{(1)}_k = \ket{k} \bra{k} $. The probability to get a state
$\ket {k} $ after projection is
\begin{align}
p_k = \frac{1}{N}\braket{\Phi|\Pi^{(1)}_k|\Phi}
\end{align}
The reduced density matrix is given by
\begin{align}\label{infreduced}
\rho^{(N-1)} = \sum_k p_k \ket{\phi_k} \bra{\phi_k}
\end{align}
After obtaining the reduced density matrix one can calculate the von Neumann
entropy with the standard formula
\begin{align}\label{entropy}
S[\rho^{(N-1)}]= -\text{Tr} \rl{\rho^{(N-1)} \ln \rho^{(N-1)}} = -\sum_i \lambda_i 
\ln \lambda_i
\end{align}
where $\lambda_i$ is an eigenvalue of the reduced density matrix.

\subsection{Second Quantization}
The above method can be recast in the language of second quantization which is 
useful to generalize the method for studying anyons. Let $ a_{k} $ be the 
annihilation operator such that $ a^\dagger_{k} \ket{0} = \ket{k}$. The inner 
product between the single particle state $ \ket{k} $ and the $ N $-particle 
state $ \ket{\Phi} $ is 
\begin{align}
a_{k} \ket{\Phi} \equiv \bra{k} \cdot \ket{\Phi}
\end{align}
Using the above expression, the formula for the reduced density 
matrix in the second quantization formalism is
\begin{align}\label{infred}
\rho^{(N-1)} \rl{\Phi}  = \frac{1}{\mathcal{N}}\sum_k a_{k}\ket{\Phi}\bra{\Phi} 
a^\dagger_{k} 
\end{align}
where $\mathcal{N}= \braket{\Phi|\sum_k a^\dagger_{k}a_{k}|\Phi} $.

\subsection{Local and Nonlocal Measurement}
In composite systems of distinguishable particles, entanglement is identified 
by performing local measurements on individual particles. In the case of 
indistinguishable particles, since it is not possible to address individual 
particles, a suitable definition of local one-particle measurement is required.
Compagno and Lo Franco \cite{franco2016quantum} give the following definition of local 
one-particle measurement: \\\\
\textbf{Definition.}  \textit{A local one-particle measurement for systems of 
identical particles is the measurement of a property of one particle performed 
on a localized region of space M (site or spatial mode) where the particle has 
nonzero probability of being found.}\\\\
The entanglement, determined by local measurements according to the above 
definition, can be quantified via the entropy of the reduced 
density matrix obtained by the localized partial trace. Here, the localized 
partial trace, in the language of second quantization, is obtained by equation 
(\ref{infred}), with the sum over the index $ k $ limited to the subset 
$ k_M $ corresponding to the subspace $ \mathcal{B}^{(1)}_M $ of one-particle 
basis states localized in $ M $:
\begin{align}
\rho^{(N-1)}_M \rl{\Phi}  = \frac{1 }{\mathcal{N}_M} \sum_{k \in \mathcal{B}^{(1)}_M } a_{k} \ket{\Phi}\bra{\Phi} a^\dagger_{k}
\end{align}
where $ \mathcal{N}_M = \braket{\Phi|\sum_{k\in\mathcal{B}^{(1)}_M} 
a^\dagger_{k}a_{k}|\Phi}$. The entropy is then found by the usual formula
\begin{align}
\begin{aligned}
E_M(\Phi) = S[\rho^{(N-1)}_M \rl{\Phi} ] = -\text{Tr} \rl{\rho^{(N-1)}_M \rl{\Phi}
 \log \rl{ \rho^{(N-1)}_M \rl{\Phi} }}
\end{aligned}
\end{align}
A nonlocal measurement, on the other hand, consists in performing the partial 
trace using a non-local one-particle basis $ \mathcal{B}^{(1)}_{M'}  $, where 
$ M' $ is a non-local spatial region. The entanglement can be quantified using 
the entropy, as before, but the corresponding reduced density matrix is 
obtained by performing the partial trace using the non-local one-particle 
basis in $M'$. It is possible that even though the local 
measurements performed on certain states of the system indicate that the 
particles are not entangled, an `induced entanglement' can be observed if we 
perform a non-local measurement on the same state. This fact confirms that 
the entanglement of indistinguishable particles is sensitive to the non-local 
character of the measurement. In the present work we will explore both types 
of entanglement by performing partial traces using local as well as non-local 
one-particle bases in the model we are studying.

\subsection{Basis Transformations and Partial Trace}

An important property of the above approach, not sufficiently emphasised in
the original paper, but noted elsewhere 
\cite{benattiEntanglementIndistinguishableParticle2020}, is as follows: the 
above definition of the reduced density matrix is independent of 
the choice of the single-particle basis used in the partial trace 
computation.\\ 

\noindent Proof:  Consider a unitary transformation $U$ in the one-particle 
subspace $ \mathcal{B}^{(1)}_M $ and the basis transformation
\begin{align}
a^\dagger_{k} = \sum_{\tilde{k}} U_{k \tilde{k}} b^\dagger_{\tilde{k}}, \quad  
\ket{k} \in \mathcal{B}^{(1)}_M 
\end{align}
Corresponding to the one-particle basis states $\ket{k}\notin
\mathcal{B}^{(1)}_M $, we may choose $ a^\dagger_{k} = b^\dagger_{\tilde{k}}$.
One can then show 
\begin{align}
\rho^{(N-1)}_M \rl{\Phi} = \frac{1 }{\mathcal{N}_M} \sum_{k \in 
\mathcal{B}^{(1)}_M } a_{k} \ket{\Phi}\bra{\Phi} a^\dagger_{k}
= \frac{1 }{\mathcal{N}_M} \sum_{\tilde{k} \in \mathcal{B}^{(1)}_M } 
b_{\tilde{k}} \ket{\Phi}\bra{\Phi} b^\dagger_{\tilde{k}}
\end{align}
where $\mathcal{N}_M = \braket{\Phi|\sum_{k\in\mathcal{B}^{(1)}_M } 
a^\dagger_{k}a_{k}|\Phi} =\braket{\Phi|\sum_{\tilde{k} \in 
\mathcal{B}^{(1)}_M } b^\dagger_{\tilde{k}}b_{\tilde{k}}|\Phi}$. The 
details are given in Appendix \ref{basistrans}.

\subsection{Singular Value Decomposition}

To get further insight into the Compagno-Lo Franco method, we can try to
recast it in a form reminiscent of the Schmidt decomposition for bipartite
systems of distinguishable particles. As may be expected, we use the
singular value decomposition and derive an alternative expression for
the von Neumann entropy.

Consider a system of two distinguishable particles to begin with.  The total 
Hilbert space is $ H_A\otimes H_B $. A two particle pure state $ \ket{\phi} $ 
in this space can be expanded as $\ket{\phi} = \sum_{i,j} \pmb{\phi}_{i,j}
\ket{i_A} \otimes \ket{j_B} $ where $ \lbrace \ket{i_A} \rbrace $ is an 
orthonormal basis of $ H_A  $ and $ \lbrace \ket{j_B} \rbrace $ is an 
orthonormal basis of $ H_B  $. The reduced density matrix of the particle 
$ A $ is given by
\begin{align}
\begin{aligned}
\rho_A = \text{Tr}_B (\ket{\phi}\bra{\phi}  ) = \sum_{j} \bra{j_B} \cdot 
\ket{\phi}\bra{\phi} \cdot \ket{j_B}
\end{aligned}
\end{align}
In terms of the coefficient matrix $ \pmb{\phi} $, the reduced density matrix 
of the particle $ A $ is $ \rho_A = \pmb{\phi} \pmb{\phi}^\dagger $. 
Similarly, the reduced density matrix of the particle $ B $ is $ \rho_B = 
\rl{\pmb{\phi}^\dagger \pmb{\phi}}^* $. The von Neumann entropy of the reduced 
density matrix $  \rho_A $ is given by
\begin{align}
\begin{aligned}
S[\rho_A ] = -\text{Tr} \rl{ \rho_A \log \rho_A}  = -\sum_{i} \lambda^{(A)}_i 
\log \lambda^{(A)}_i
\end{aligned}
\end{align}
where $ \lambda^{(A)}_i $ is an eigenvalue of the reduced density matrix 
$ \rho_A $. 

Suppose we perform the singular value decomposition of the matrix $\pmb{\phi}$ 
as $ \pmb{\phi} = U D V^\dagger $, where $ U $ and $ V $ are unitary matrices 
and $ D $ is a diagonal matrix. Then
\begin{align}
\begin{aligned}
\rho_A = \pmb{\phi} \pmb{\phi}^\dagger  =U D V^\dagger V D^\dagger 
U^\dagger = U D D^\dagger U^\dagger.
\end{aligned}
\end{align}
From the above relation, it is clear that $ U $ diagonalizes $ \rho_A  $ and 
the eigenvalues of $ \rho_A $ are related to the singular values $ \sigma_i $ 
of $ \pmb{\phi} $ as
\begin{align}
\lambda^{(A)}_i = |\sigma_i|^2
\end{align}
This gives an expression for entropy in terms of the singular values of 
$ \pmb{\phi}$
\begin{align}
S[\rho_A ]= -\sum_i |\sigma_i|^2 \log (|\sigma_i|^2)
\end{align}
With a similar argument we can also show that $ S[\rho_B ]= -\sum_i 
|\sigma_i|^2 \log (|\sigma_i|^2) $ and conclude $ S[\rho_A ] =  S[\rho_B ]$. 

In the case of distinguishable particles, the singular value decomposition 
helps to perform the Schmidt decomposition of a pure state and measure 
entanglement using the Schmidt rank. The method discussed above relates the 
von Neumann entropy to the Schmidt coefficients 
\cite{nielsenQuantumComputationQuantum2002}. However, the same definition of 
Schmidt decomposition cannot be used in the case of indistinguishable particles
as the total Hilbert space of the system is not a product of Hilbert spaces 
of the constituent systems. But, the von Neumann entropy can still be related 
to the singular values of the coefficient matrix $ \pmb{\phi} $, as we will 
show next.

In the Compagno - Lo Franco method, the definition of the reduced 
density matrix obtained by a partial trace in the basis of the one-particle 
subspace $ \mathcal{B}^{(1)}_M $ is
\begin{align}
\rho^{(1)}_M \rl{\Phi^{(2)}}  = \frac{1 }{\mathcal{N}_M} \sum_{k\in\mathcal{B}^{(1)}_M } a_{k} \ket{\Phi^{(2)}}\bra{\Phi^{(2)}} a^\dagger_{k}
\end{align}
where $ \ket{\Phi^{(2)}} $ is a two-particle state.

Suppose we consider the matrix elements of the reduced density matrix 
$ \rho^{(1)}_M \rl{\Phi^{(2)}}   $ which are given by
\begin{align}
\begin{aligned}
\rl{ \rho^{(1)}_M \rl{\Phi^{(2)}} }_{m,n} &= \braket{m| \rho^{(1)}_M \rl{\Phi^{(2)}} |n}\\
&=\frac{1 }{\mathcal{N}_M} \sum_{k \in \mathcal{B}^{(1)}_M }\bra{0} a_m a_{k} 
\ket{\Phi^{(2)}}\bra{\Phi^{(2)}} a^\dagger_{k} a^\dagger_n \ket{0}\\
&= \sum_{k\in \mathcal{B}^{(1)}_M}\pmb{\varphi}_{mk} \pmb{\varphi}_{nk}^*  
\qquad 
\end{aligned}
\end{align}
where $\pmb{\varphi}_{mk} = \frac{\bra{0}a_m a_k \ket{\Phi^{(2)}}}{\sqrt
{\mathcal{N}_M}}$, $\ket{m}$ and $\ket{n} $ are one-particle basis states 
and the indices $ m,n=1,..,d $ where $ d $ is the dimension of the 
single-particle Hilbert space. In terms of the matrix $\pmb{\varphi}$ the reduced density matrix is 
\begin{align}
\rho^{(1)}_M \rl{\Phi^{(2)}} = \pmb{\varphi} \pmb{\varphi}^\dagger
\end{align}
Note that since $k\leq d$, $\pmb{\varphi} $ is a rectangular matrix in general. 

Similar to the case of distinguishable particles, one may perform the singular
value decomposition of $\pmb{\varphi}$ and obtain the eigenvalues of the 
reduced density matrix as $ \lambda_i = |\sigma_i|^2 $, where $ \sigma_i $ are 
the singular value of the matrix $ \pmb{\varphi} $. This gives the 
expression for entropy in terms of the singular values of $ \pmb{\varphi}$
\begin{align}
E_M(\Phi^{(2)}) = S[\rho^{(1)}_M \rl{\Phi^{(2)}} ]= -\sum_i |\sigma_i|^2 \log (|\sigma_i|^2)
\end{align}
Thus, with the knowledge of singular values of $ \pmb{\varphi} $ we can 
calculate the entropy. This approach may be useful while computing eigenvalues 
numerically where the singular value decompositiom may be superior to 
eigenvalue decomposition. However, for our purposes we will use the original 
definition to compute the entropy.

\section{The Anyonic Hubbard Model}\label{theahm}
The Hubbard Hamiltonian \cite{hubbardElectronCorrelationsNarrow1963}
describes the physics of electronic correlations in narrow energy bands.
The model has been extensively studied in low dimensions; exact solutions
of the model in higher dimensions are not known
\cite{esslerOneDimensionalHubbardModel2005}.

In its general form, the the model is described  by the Hamiltonian
\begin{align}
\begin{aligned}
H_{F} = -\sum_{i,j,s} \kappa_{i,j} f_{i,s}^\dagger f_{j,s} +\sum_{i,j,k,l,s,s'} U_{i,j,k,l} f_{i,s}^\dagger f_{j,s'}^\dagger f_{k,s'} f_{l,s'} -\mu  \sum_{i,s} f^\dagger_{i,s} f_{i,s}
\end{aligned}
\end{align}
where $ s,s'= \uparrow, \downarrow$ denote the spin and $ i,j,k,l = 1,2,..,L $
denote the lattice sites. $ f_{j,s}^\dagger ( f_{j,s}) $ denotes a fermion
creation(annihilation) operator creating(annihilating) a particle with spin $s$
at a lattice site $j$. $\mu$ is the chemical potential. The hopping matrix
elements are given by
\begin{align}
\begin{aligned}
\kappa_{i,j} = \int d^3 x \phi^*(x-R_i) \rl{h_1}  \phi(x-R_j) 
\end{aligned}
\end{align}
and the interaction parameters are
\begin{align}
\begin{aligned}
U_{i,j,k,l} = \int d^3 x d^3y \phi^*(x-R_i) \phi^*(y-R_j) \rl{U(x,y)}  \phi(y-R_k) \phi(x-R_l)
\end{aligned}
\end{align}
where $ h_1 $ is the one-particle Hamiltonian and $ U(x,y) $ is the two-body
interaction \cite{esslerOneDimensionalHubbardModel2005}. $R$ denotes a lattice
vector and $\phi(x-R)$ the Wannier function.

The original Hubbard model assumes that the off-site interactions are
negligible compared to the onsite interaction $U_{i,i,i,i}$ and also retains
only the nearest neighbor hopping matrix elements. For our purposes, we
find it useful to retain the nearest neighbor off-site interactions and
work with the Hamiltonian
\begin{align}
\begin{aligned}
H_{F} =- \kappa \sum_{\braket{i,j}} \sum_{s} a^\dagger_{i,s} a_{j,s} + 
U\sum_{i}n_{i\uparrow} n_{i\downarrow}+ V \sum_{\braket{i,j}}\sum_{s,s'} 
n_{i,s} n_{j,s'}+J \sum_{\braket{i,j}} \sum_{s,s'} a^\dagger_{i,s} 
a^\dagger_{j,s'} a_{i,s'}a_{j,s} - \mu \sum_{i,s} n_{i,s}
\end{aligned}
\end{align}
where $ \kappa, U, $ and $ \mu $ denote the hopping parameter, the strength of
the on-site interaction, and the chemical potential, respectively. $V$ and
$J$ denote nearest-neighbour off-site interactions. Note that all the
parameters in this model assume real values. Also, the number operator is
given by the usual expression, $ n_{i,s} = a^\dagger_{i,s} a_{i,s} $.
The anyon creation ($ a^\dagger_{j,s} $) and annihilation ($ a_{j,s} $)
operators satisfy the following algebra

\begin{align}\label{comm}
\begin{aligned}
a_{j,s} a_{k,s'} +e^{i \nu \sgn(j-k)} a_{k,s'} a_{j,s}  &=0\\
a_{j,s} a_{k,s'}^\dagger +e^{-i \nu \sgn(j-k)} a_{k,s'}^\dagger a_{j,s} &=\delta_{j,k} \delta_{s,s'}
\end{aligned}
\end{align}
where the sign function $\sgn(x) = -1$ when $ x<0 $, $ \sgn(x) = 0$ when
$ x=0$, and $ \sgn(x) = 1 $ when $ x>0 $.
This algebra becomes fermionic when $\nu=0$, and represents hardcore bosons,
which behave like as bosons offsite and fermions onsite, in the limit
$\nu =\pi$.

\subsection{The Two-site Anyonic Hubbard Model: Exact diagonalization}

The Hamiltonian of the two-site Anyonic Hubbard model is given by
\begin{align}
\begin{aligned}
H =- \kappa \sum_{\braket{i,j}} \sum_{s}  a^\dagger_{i,s} a_{j,s} + U \sum_{i}n_{i\uparrow} n_{i\downarrow}+ V \sum_{\braket{i,j}}\sum_{s,s'} n_{i,s} n_{j,s'}+J \sum_{\braket{i,j}} \sum_{s,s'} a^\dagger_{i,s} a^\dagger_{j,s'} a_{i,s'}a_{j,s} - \mu \sum_{i,s} n_{i,s}
\end{aligned}
\end{align}
where $ i,j=1,2 $.

Note that the Hamiltonian commutes with the total number operator
$ N = \sum_{i,s} n_{i,s} $ and the total spin in the $ Z $-direction
$ S_z = \sum_{i} \rl{n_{i,\uparrow} - n_{i,\downarrow}} $; these are
conserved quantities. Hence, the structure of the Hamiltonian is block
diagonal where each block corresponds to fixed values of $ N $ and $S_z$.
Accordingly, we choose the following basis for the Fock space :
\begin{align}
\begin{aligned}
\{\ket{0},\ket{1\uparrow},\ket{2\uparrow},\ket{1\downarrow},\ket{2\downarrow},\ket{1\uparrow,2\uparrow},\ket{1\uparrow,1\downarrow}, \ket{2\uparrow,2\downarrow},\ket{1\uparrow,2\downarrow},\ket{1\downarrow,2\uparrow},\\ \ket{1\downarrow,2\downarrow},\ket{1\uparrow,1\downarrow,2\uparrow},\ket{1\uparrow,2\uparrow,2\downarrow},\ket{1\uparrow,1\downarrow,2\downarrow},\ket{1\downarrow,2\uparrow,2\downarrow},\ket{1\uparrow,1\downarrow,2\uparrow,2\downarrow} \} 
\end{aligned}
\end{align}
where, in general, $ \ket{j_1s_1,j_2s_2,..,j_{N}s_{N}}  = a^\dagger_{j_1,s_1}a^
\dagger_{j_2,s_2}...a^\dagger_{j_N,s_N} \ket{0}$.
In this basis the Hamiltonian can be written in the block diagonal form
\begin{align}
\begin{aligned}
H = H_{0,0} \oplus H_{1,\frac{1}{2}}\oplus H_{1,-\frac{1}{2}}\oplus H_{2,1}
\oplus H_{2,0}\oplus H_{2,-1}\oplus H_{3,\frac{1}{2}}\oplus H_{3,-\frac{1}{2}}
 \oplus H_{4,0}
\end{aligned}
\end{align}
where
\begin{align}
\begin{aligned}
H_{0,0} &= [0]\\
H_{1,\frac{1}{2}} &= H_{1,-\frac{1}{2}} =
\begin{bmatrix}
-\mu  & -\kappa  \\
-\kappa  & -\mu  \\
\end{bmatrix} \\
H_{2,1} &=H_{2,-1}=\begin{bmatrix}
2 (V-J \cos (\nu )-\mu)
\end{bmatrix} \\
H_{2,0} &= \begin{bmatrix}
U-2 \mu  & 0 & -\kappa  & \kappa  \\
0 & U-2 \mu  &   -\kappa e^{i \nu }
& \kappa  e^{i \nu } \\
-\kappa  & -\kappa e^{-i \nu } & 2
(V-\mu ) & -2 J \cos (\nu ) \\
\kappa  & \kappa  e^{-i \nu } & -2 J \cos (\nu )
& 2 (V-\mu ) \\
\end{bmatrix}\\
H_{3,\frac{1}{2}}&= H_{3,-\frac{1}{2}} =\begin{bmatrix}
U+4 V -2 J \cos (\nu )-3 \mu & \kappa  e^{-i
        \nu } \\
\kappa  e^{i \nu } & U+4
V-2 J \cos (\nu )-3 \mu \\
\end{bmatrix} \\ H_{4,0} &=
\begin{bmatrix}
2 (U+4 V-2 J \cos (\nu )-2 \mu ) \\
\end{bmatrix}
\end{aligned}
\end{align}
Four eigenvalues of the Hamiltonian are readily obtained by reading them
off from $ 1 \times 1 $ blocks of the matrix. The rest of them can be found by
solving the $ 2 \times 2 $ blocks and the $ 4\times 4 $ block. The eigenvalues
and eigenvectors of the Hamiltonian are listed in the Appendix \ref{eigen}.

\subsection{The Anyonic Hubbard Model in External Fields}\label{extfield}
Applying an external field to the system allows us to control certain
physical properties and thus the behavior of the system. In our case, we apply 
an external electric field to the system which modifies the two-site 
Hamiltonian.

\begin{figure}[H]
\centering
\begin{tikzpicture}
\draw (-1,0) --(6,0);
\draw[fill] (0,0) circle (3pt);
\draw[fill] (1,0) circle (3pt);
\draw[fill] (2,0) circle (3pt);
\draw[fill] (3,0) circle (3pt);
\draw[fill] (4,0) circle (3pt);
\draw[fill] (5,0) circle (3pt);
\draw[->] (4.65,0.5)--(5,0.5);
\draw[<-] (4,0.5)--(4.35,0.5);
\node at (4.5,0.5) {$ a $};
\node[below] at (2.5,-2) {O};
\draw[->] (2.5,-2) -- (2.05,-0.2);
\draw[->] (2.5,-2) -- (2.95,-0.2);
\node at (2,-1) {$ \mathbf{R}_j $};
\node[right] at (2.8,-1) {$ \mathbf{R}_{j+1} $};
\end{tikzpicture}
\caption{The one-dimensional anyonic Hubbard model on a chain. $ a $ is the lattice spacing.}
\label{diagram}
\end{figure}
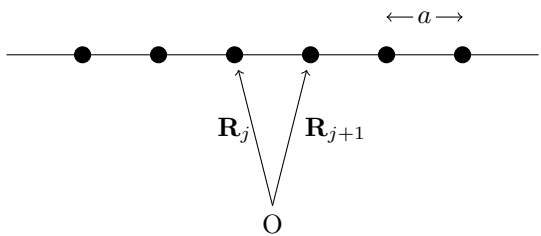

Let $\mathbf{R}_j$ and $\mathbf{R}_{j+1}$ be the position coordinates of the 
the $j$th and the $(j+1)$th site respectively with respect to a fiducial 
reference point $O$ (Figure \ref{diagram}). A particle hopping from one site to another site 
corresponds to the creation of an anti-particle at the former site, and a 
particle at the latter site. In the presence of an external field, the two 
sites should then be linked by a flux (Wilson) line $e^{i\lambda_{j, j+1}}$ 
joining the two sites, where 
\begin{equation}
\lambda_{j,j+1} = {e\over c}\int_{\mathbf{R}_j}^{\mathbf{R}_{j+1}} dx_\alpha A^\alpha(\vec x, t)  
\end{equation}

We choose the vector potential to be $\pmb{A}(t) = \pmb{\Omega}t\theta(t)$, 
where $ \theta(t) $ is the unit step function and choose the gauge where the
scalar potential $ \Phi =0$. The electric field is given by $ \pmb{\Omega} 
= -d\pmb{A}(t)/dt$. With the inclusion of the Wilson line in the Hamiltonian,
the hopping parameter becomes time-dependent and complex. However, the other
interaction parameters in the Hamiltonian do not change. The modified hopping
parameter is given by $\kappa(t) = -\kappa e^{-i \pmb{A}(t). 
(\pmb{R}_1-\pmb{R}_2)} =  -\kappa e^{i \Omega t}$ if we assume the first 
lattice site to be to the left of second lattice site and $ \kappa $ is real
\cite{balasubramanianExactTimeEvolution2017}. The modified Hamiltonian is
\begin{align}
\begin{aligned}
H(t) &=- \kappa  \sum_{s}  \rl{e^{i \Omega t}a^\dagger_{1,s} a_{2,s}+e^{-i 
\Omega t} a^\dagger_{2,s} a_{1,s}} + U \sum_{i}n_{i\uparrow} n_{i\downarrow}\\
&+ V \sum_{\braket{i,j}}\sum_{s,s'} n_{i,s} n_{j,s'}+J \sum_{\braket{i,j}} 
\sum_{s,s'} a^\dagger_{i,s} a^\dagger_{j,s'} a_{i,s'}a_{j,s} - \mu \sum_{i,s} 
n_{i,s}
\end{aligned}
\end{align}
where $ i,j=1,2 $.
We note that if we write the Hamiltonian in the block diagonal form, only the
blocks containing the hopping parameter are modified. Consider the block
\begin{align}
\begin{aligned}
 H_{2,0}(t)  &=\left[
 \begin{array}{cccc}
 U-2 \mu  & 0 &  -\kappa e^{-i (\nu- \Omega t)} & \kappa  e^{-i (\nu-\Omega t)} \\
 0 & U-2 \mu  &  - \kappa e^{-i \Omega t
 }& \kappa  e^{-i \Omega t } \\
 -  \kappa e^{i (\nu -\Omega t )} &
 - \kappa e^{i \Omega t }& 2
(V-\mu ) & -2 J \cos (\nu ) \\
 \kappa  e^{i (\nu -\Omega t )} & \kappa  e^{i 
        \Omega t } & -2 J \cos (\nu ) & 2 (V-\mu ) \\
 \end{array}
 \right]
\end{aligned}
\end{align}
If we perform a unitary transformation by the unitary operator $ W_1  = 
\text{diag} \lbrace e^{- i \Omega t}, e^{i \Omega t},1,1\rbrace $ the
Hamiltonian is converted to $ \tilde{H}_{2,0} = W_1^\dagger H_{2,0}(t) W_1 
- i W_1^\dagger \dot{W}_1 $, which in the matrix form is
\begin{align}
\begin{aligned}
\tilde{H}_{2,0} =\left[
\begin{array}{cccc}
U-2 \mu +\Omega  & 0 &  -\kappa e^{-i \nu
} & \kappa  e^{-i \nu } \\
0 & U-2 \mu -\Omega  & -\kappa  & \kappa  \\
  -\kappa e^{i \nu } & -\kappa  & 2
(V-\mu ) & -2 J \cos (\nu ) \\
\kappa  e^{i \nu } & \kappa  & -2 J \cos (\nu )
& 2 (V-\mu ) \\
\end{array}
\right]
\end{aligned}
\end{align}
Since $ \tilde{H}_{2,0} $ is time-independent, the time evolution operator
can be written down as $ W_2 = e^{-i \tilde{H}_{2,0} t} $. (We set $ \hbar=1 $ for convenience.) Therefore, the
total time evolution operator is given by $ \tilde{W}= W_1 W_2 $. Similarly
the unitary operators corresponding to the $ 2 \times 2 $ blocks

\begin{align}
\begin{aligned}
H_{1,\frac{1}{2}} &= H_{1,-\frac{1}{2}} =
\begin{bmatrix}
-\mu  & -\kappa e^{i \Omega t}  \\
-\kappa e^{-i \Omega t}  & -\mu  \\
\end{bmatrix} \\, \quad H_{3,\frac{1}{2}}&= H_{3,-\frac{1}{2}} =\begin{bmatrix}
U+4 V-2 J \cos (\nu ) -3 \mu & \kappa  e^{-i
        \rl{\nu-\Omega t} } \\
\kappa  e^{i \rl{\nu-\Omega t }} & U+4
V-2 J \cos (\nu )  -3 \mu \\
\end{bmatrix} 
\end{aligned}
\end{align}
can be found by performing a unitary transform with the matrix $ W_3 =$ diag
$ \lbrace 1, e^{-i \Omega t}\rbrace  $. This will give the following
transformed Hamiltonians

\begin{align}
\begin{aligned}
\tilde{H}_{1,\frac{1}{2}} &= \tilde{H}_{1,-\frac{1}{2}} =
\begin{bmatrix}
-\mu  & -\kappa  \\
-\kappa   & -\mu -\Omega  \\
\end{bmatrix} \\,\quad\tilde{H}_{3,\frac{1}{2}}&= \tilde{H}_{3,-\frac{1}{2}} =
\begin{bmatrix}
U+4 V-2 J \cos (\nu )  -3 \mu & \kappa  e^{-i
        \nu  } \\
\kappa  e^{i \nu } &U+4
V  -2 J \cos (\nu )  -3 \mu -\Omega \\
\end{bmatrix} 
\end{aligned}
\end{align}
Again the transformed Hamiltonians are time-independent and the time evolution
operators can be found as before. Finally, the total time evolution operator
$ W(t) $ can be constructed by taking a direct sum of the time evolution
operators corresponding to each block.

\section{Entanglement Entropy} \label{ente}
In this section, we study the entanglement entropy in the two-site Anyonic
Hubbard Model. We consider two-particle states and find the corresponding reduced density matrix and entropy using the information
theoretic approach oulined earlier. Since the exact eigenstates of the
Hamiltonian are known in this case, we study the entropy corresponding to
the following two-particle eigenstates
\begin{align}
\begin{aligned}
\ket{\phi_{2,1}}&=\ket{1\uparrow,2\uparrow}\\\ket{\phi_{2,2}}&= -e^{-i \nu }
\ket{1\uparrow,1\downarrow}+\ket{2\uparrow,2\downarrow}\\  \ket{\phi_{2,3}}&=
\ket{1\uparrow,2\downarrow}+\ket{1\downarrow,2\uparrow}\\  \ket{\phi_{2,4}}&=
\frac{4 \kappa e^{-i \nu }}{\Upsilon- v_2}\ket{1\uparrow,1\downarrow}+\frac{4 
\kappa  }{ \Upsilon-v_2}\ket{2\uparrow,2\downarrow}-\ket{1\uparrow,2\downarrow}
+\ket{1\downarrow,2\uparrow} \\  \ket{\phi_{2,5}}&=-\frac{4 \kappa e^{-i \nu }}
{\Upsilon+ v_2}\ket{1\uparrow,1\downarrow}-\frac{4 \kappa  }{\Upsilon+ v_2}\ket
{2\uparrow,2\downarrow}-\ket{1\uparrow,2\downarrow}+\ket{1\downarrow,2\uparrow}
\\  \ket{\phi_{2,6}}&= \ket{1\downarrow,2\downarrow}\\
\end{aligned}
\end{align}
where, $\Upsilon = \sqrt{\left(16 \kappa ^2+v_2^2\right)}$ and $v_2= -2 J \cos
(\nu )+U-2 V$.

The entanglement, as discussed before, can be determined using local as well 
as non-local measurements. In the case of the two-site Anyonic Hubbard Model, 
the local projective measurements correspond to measuring properties of the 
particles at the first or the second lattice site, while a non-local 
measurement involves both lattice sites. Accordingly, we choose the local 
one-particle bases $\mathcal{B}_1 = \lbrace\ket{1\uparrow},\ket{1\downarrow} 
\rbrace$ and $\mathcal{B}_2 = \lbrace\ket{2\uparrow},\ket{2\downarrow}\rbrace$ 
to perform localized partial traces at site 1 and site 2, respectively, while 
the non-local basis $\mathcal{B}_{12} = \lbrace\ket{1\uparrow}, 
\ket{1\downarrow}, \ket{2\uparrow}, \ket{2\downarrow}\rbrace$ is chosen to 
perform the non-local partial trace.

The list of reduced density matrices after performing the partial 
trace onto various one-particle subspaces is given in the Appendix 
\ref{reducedmatrices}. The corresponding entropy can be found by the standard 
formula \ref{entropy}.

In the case of local projective measurements at site 1, we observe that the
entropy $E_1\rl{\phi_{2,1}} =0$ and $ E_1\rl{\phi_{2,6}} =0$. That is, 
particles in the states $ \ket{\phi_{2,1}} $ and $ \ket{\phi_{2,6}} $ are 
not entangled. However, the entropy is equal to unity for all the  other 
four states: $ E_1\rl{\phi_{2,2}} =E_1\rl{\phi_{2,3}} =E_1\rl{\phi_{2,4}} 
=E_1\rl{\phi_{2,5}} =1 $. Similar results are obtained corresponding a 
local projective measurement at site 2. In this case $ E_2\rl{\phi_{2,1}} 
= E_2\rl{\phi_{2,6}}=0 $ and $ E_2\rl{\phi_{2,2}} =E_2\rl{\phi_{2,3}} 
=E_2\rl{\phi_{2,4}} =E_2\rl{\phi_{2,5}} =1 $.

\begin{figure}[H]
	\centering
	\includegraphics[scale=0.7]{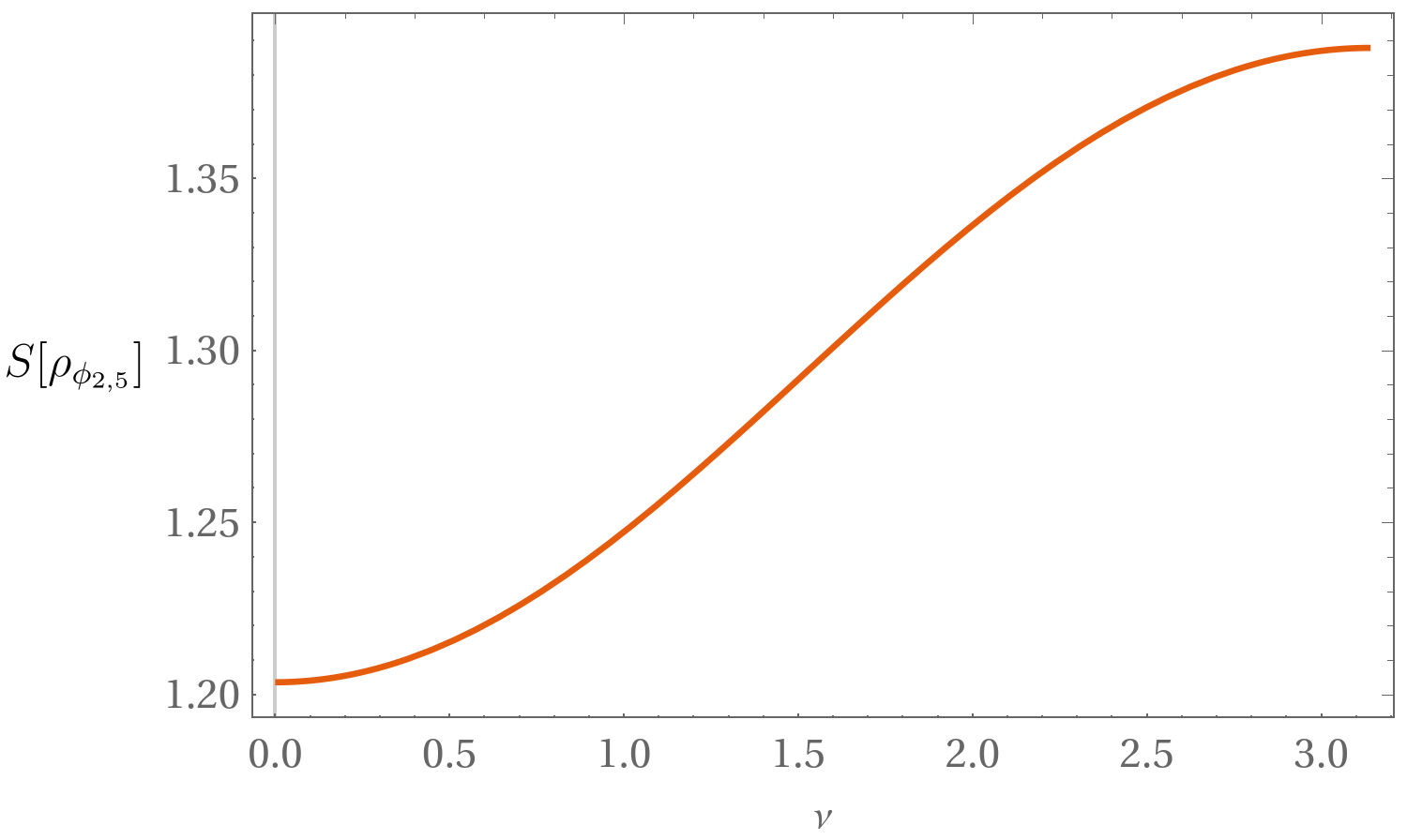} 
	\caption{Entropy $ E_{12}(\phi_{2,5}) $ is plotted against the 
statistics parameter $\nu$ ($\kappa=1,U=4,V=1,J=0.25$ and  $\mu=0.5$).}
	\label{fig:entropy}
\end{figure}

When the partial trace is made onto the non-local one-particle subspace 
$\mathcal{B}_{12} $, the values of entropy are $ E_{12}\rl{\phi_{2,1}} = 
E_{12}\rl{\phi_{2,6}}=1 $ and $ E_{12}\rl{\phi_{2,2}} = E_{12}\rl{\phi_{2,3}}
=2 $, while
\begin{align}
\begin{aligned}
E_{12}\rl{\phi_{2,4}} =E_{12}\rl{\phi_{2,5}}= -2 \rl{\frac{1}{4}-\frac{\kappa }
{\Upsilon}} \log\rl{\frac{1}{4}-\frac{\kappa }{\Upsilon}}-2 \rl{\frac{1}{4}+
\frac{\kappa }{\Upsilon}} \log\rl{\frac{1}{4}+\frac{\kappa }{\Upsilon}}
\end{aligned}
\end{align}
The entropy corresponding to the states $\ket{\phi_{2,4}}$ and $\ket
{\phi_{2,5}} $ shows dependence upon the statistical parameter. This 
dependence is plotted in Figure \ref{fig:entropy} by fixing the values of 
other interaction parameters conveniently.

\subsection{Entanglement in an External Field}\label{entexternal}

In section \ref{extfield}, we saw that introducing an external electric field
makes the Hamiltonian time-dependent. The problem can be exactly solved and 
time evolution operator found in the two-site model. This allows us to
study entropy as a function of the statistical parameter, time, and the 
external field strength.

To illustrate the above statement, we consider the non-local partial trace 
onto the one-particle subspace 
$\mathcal{B}_{12} $. We choose the initial states at time $ t=0 $ to be 
$\ket{\phi_{2,2}} $ and $ \ket{\phi_{2,5}} $. These two states can be evolved 
in time to the states $ \ket{\phi_{2,2}(t)} =W(t) \ket{\phi_{2,2}} $ and 
$ \ket{\phi_{2,5}(t)} =W(t) \ket{\phi_{2,5}}$ respectively, in the presence 
of the external electric field $ \Omega $. 

\begin{figure}
	\centering
	\includegraphics[scale=0.5]{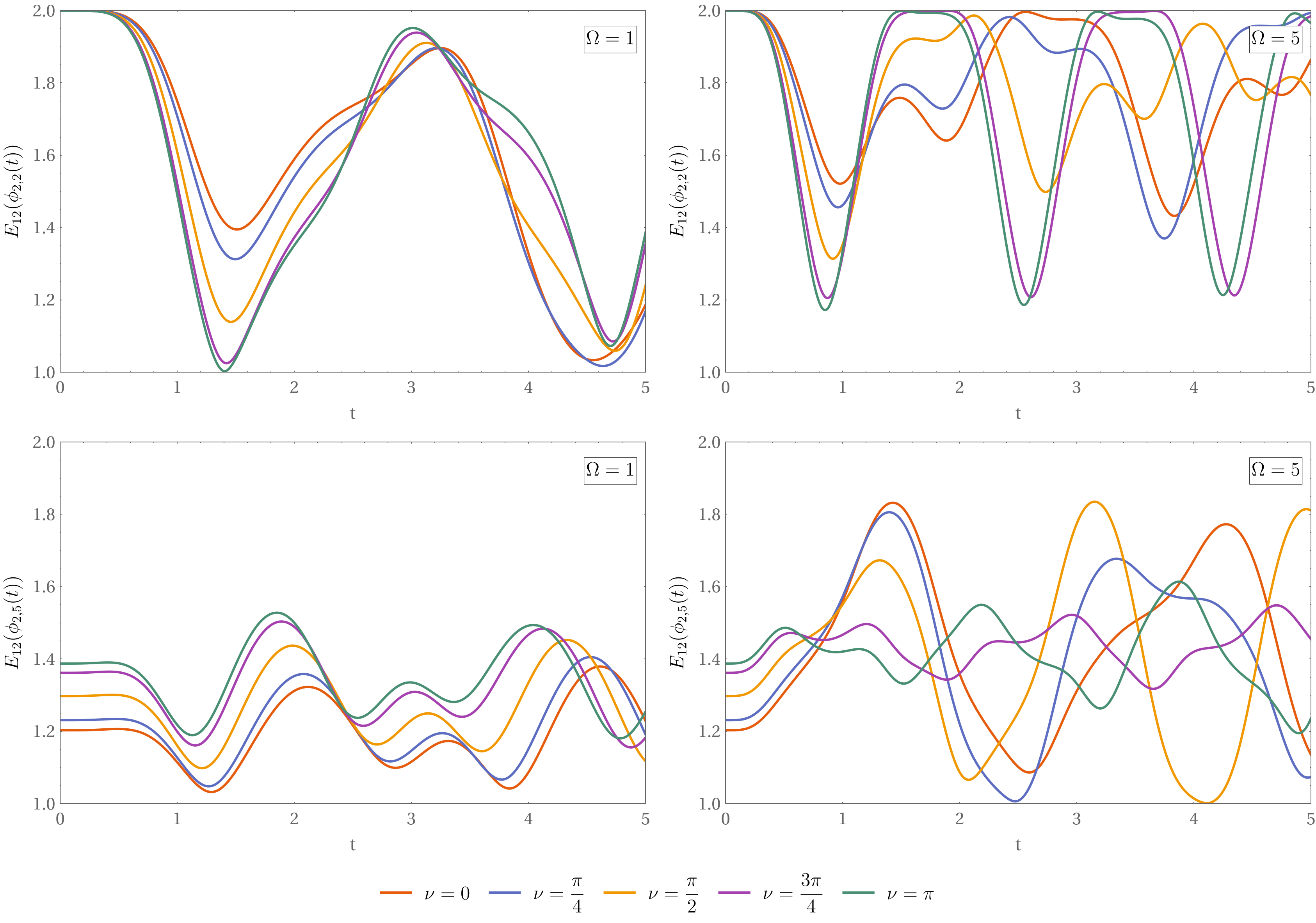}
	\caption{Plots showing time evolution of the entropy for different 
values of the statistics parameter ($\kappa=1,U=4,V=1,J=0.25,\mu=0.5$ ) and the external field strength (value 
shown in the inset) corresponding to the states $\ket{\phi_{2,2}(t)} $ and 
$  \ket{\phi_{2,5}(t)}$.}
	\label{fig:external}
\end{figure}

In Figure \ref{fig:external} we plot the variation of entropy with respect to 
time for different values of the statistics parameter and the electric field 
strength.
Note that the entropy corresponding to the initial state $\ket{\phi_{2,2}}$ is
independent of the statistics parameter when the external perturbation is 
absent. However, when the system evolves under the application of an external 
field, the entropy starts to vary as a function of the statistics parameter.

The time-variation of entropy in the presence of an external field may allow
us to control the entanglement in the system by tuning the external field.
This may have potential applications.

\subsection{Entanglement Extraction}
Extraction procedures allow one to exploit entanglement intrinsic to identical 
particle systems. A procedure to extract entanglement between the two qubits 
in the anharmonic double well was outlined in \cite{franco2016quantum}. In this 
section, will discuss the extraction procedure in the context of the two-site
Anyonic Hubbard model.

Consider, for example, a two particle state $ \ket{\Phi_i^{(2)}}  = 
\ket{1\uparrow,1\downarrow}$. In the absence of the external field, the 
entropy corresponding to the state is $E_1(\Phi_i^{(2)}) =1 $ -- it is 
maximally entangled. Since single particles cannot be individually addressed, 
this identical particle entanglement is unexploitable by local operations and 
classical communication (LOCC). 

In the presence of an external electric field $ \Omega $, however, the input 
state evolves in to a state of the form
\begin{align}\label{outstate}
\begin{aligned}
\ket{\Phi_o^{(2)}} = c_1(t) \ket{\Phi^{(2)}_{1}} +c_2(t) 
\ket{\Phi^{(2)}_{12}}+  c_3(t) \ket{\Phi^{(2)}_{2}}
\end{aligned}
\end{align}
where $ \ket{\Phi^{(2)}_{1}} = \ket{1\uparrow,1 \downarrow},
\ket{\Phi^{(2)}_{2}} = \ket{2\uparrow,2 \downarrow} $ and 
$ \ket{\Phi^{(2)}_{1 2}} =  \rl{\ket{1\uparrow,2 \downarrow} -\ket{2\uparrow,1 
\downarrow} }/\sqrt{2} $. The coefficients $ c_1(t),c_2(t) $ and $ c_3(t) $ 
depends upon the external field strength and can be tuned by adjusting  the 
external field strength and time. In Appendix \ref{coeff}, we discuss how to 
obtain these coefficients.  

The states $  \ket{\Phi^{(2)}_{1}} $ and $ \ket{\Phi^{(2)}_{2}} $ have the
same entanglement as the state $ \ket{\Phi_i^{(2)}}$, but this entanglement 
remains unexploitable by LOCC since both the particles occupy the same lattice 
sites in these states.  But, in the state $ \ket{\Phi^{(2)}_{1 2}} $, the two 
particles are at distinct sites. The state $ \ket{\Phi^{(2)}_{1 2}} $ is 
maximally entangled and the entanglement is (probabilistically) established 
between two particles at distinguishable sites. This entanglement is thus an 
exploitable resource in the LOCC paradigm.
\section{Correlation Functions}\label{corf}
In this section, we will compute the static correlation functions of anyons
in the two-site model at finite temperature. Our main interest is in
studying how the correlation functions depend upon the statistics parameter.

\begin{figure}[H]
        \centering
        \includegraphics[scale=0.5]{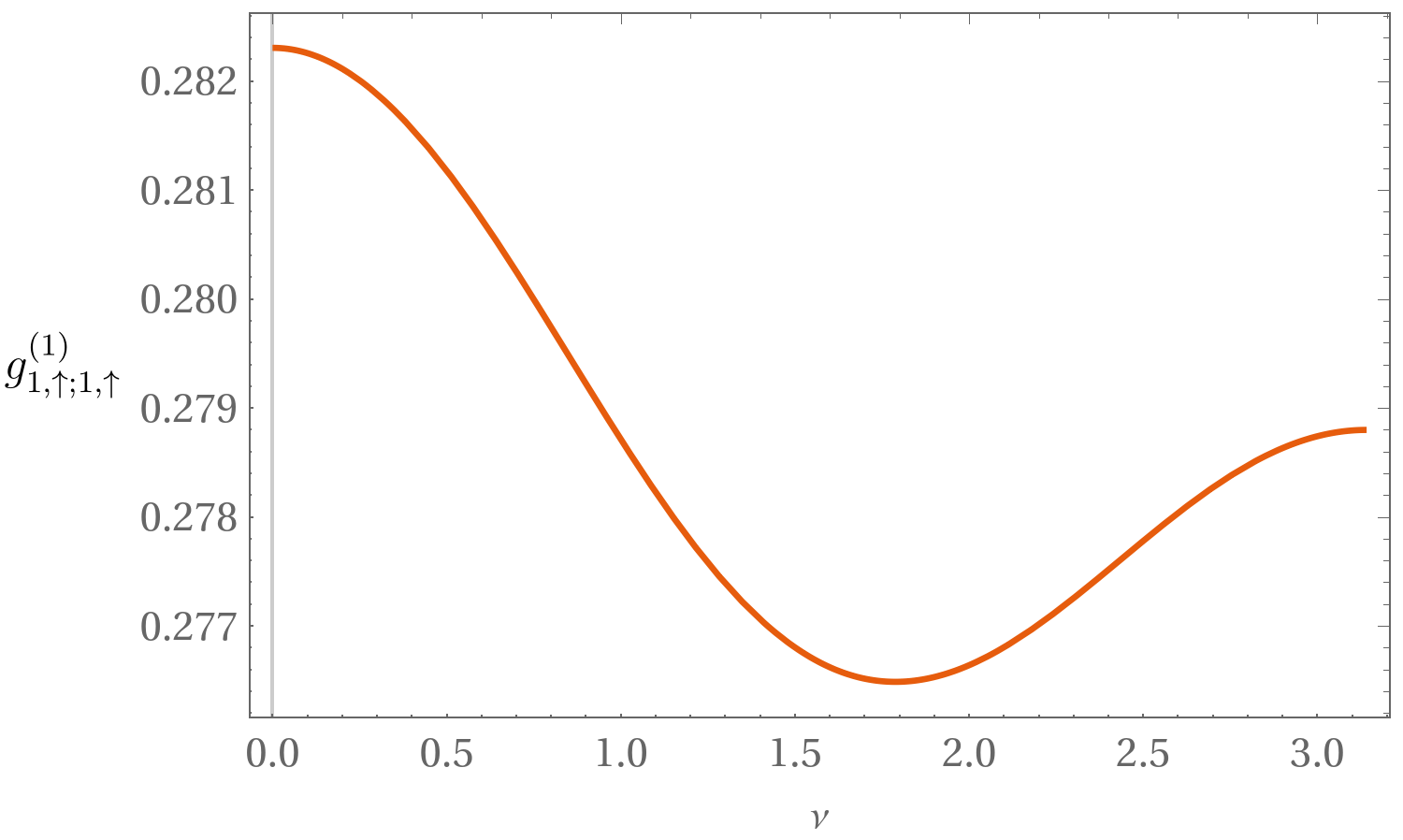}
        \includegraphics[scale=0.5]{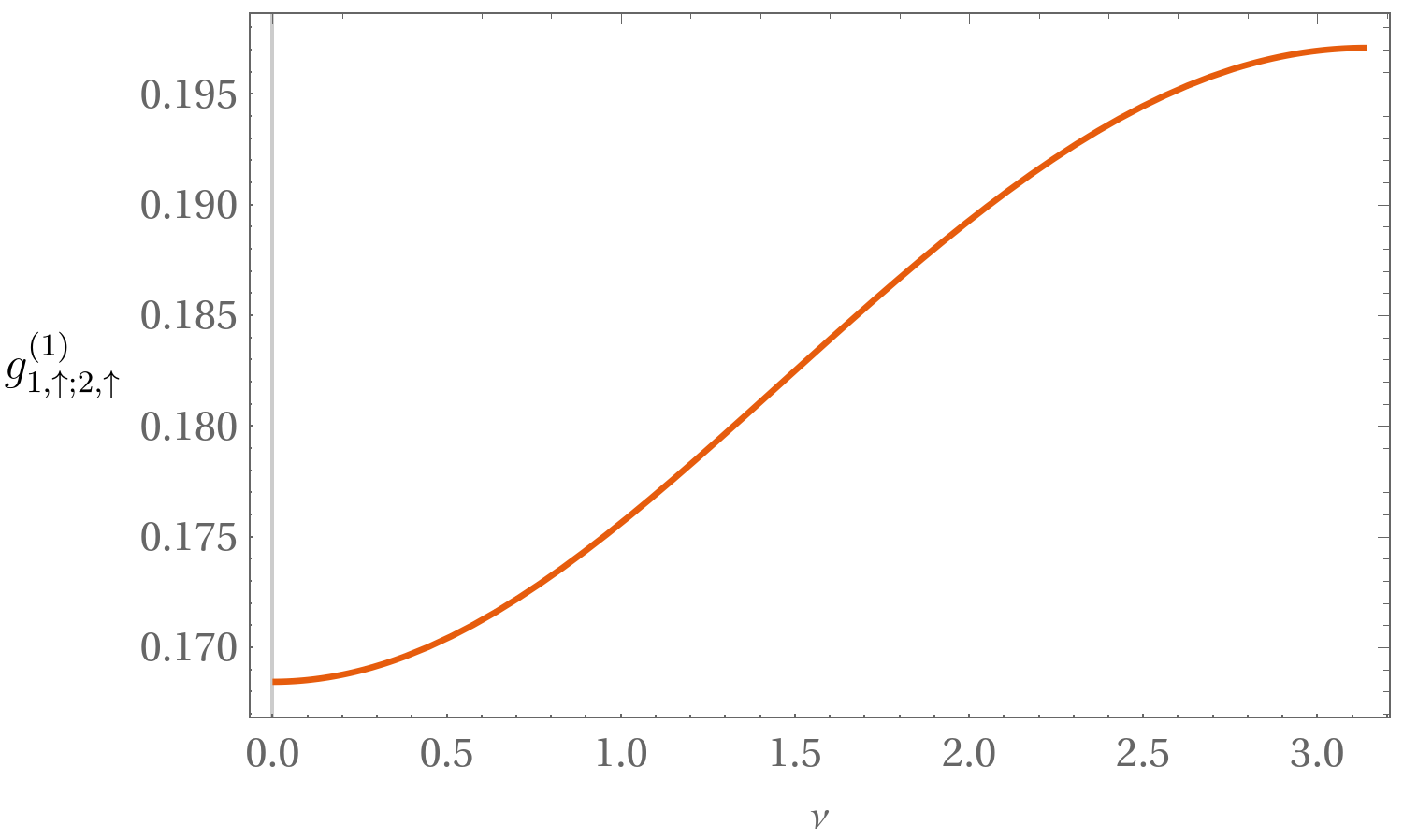}
        \caption{Plots showing dependence of two-point correlation functions on the statistical parameter ($\kappa=1,U=4,V=1,J=0.25,\mu=0.5,\beta=1 $).}
        \label{fig:twopoint}
\end{figure}

The two-point correlation function, also known as the one-particle density
matrix, measures the amplitude for removing a particle from a site $j_2$ with
spin $s_2$ and replacing it at site $j_1$ with spin $s_1$. It is given by the
formula
\begin{align}
\begin{aligned}
g^{(1)}_{j_1,s_1;j_2,s_2} =  \frac{1}{Z} \tr\rl{ a_{j_1,s_1}^\dagger a_{j_2,s_2'} e^{-\beta H} } 
\end{aligned}
\end{align}
where $ Z =\tr \rl{ e^{-\beta H} }  $ is the partition function. Since the
exact solution is known in the case of the two-site model, the correlation
function is evaluated exactly. The results are given in Appendix \ref{correlationf}. To show
the dependence on the statistics, we plot the correlation function against the
statistics parameter $ \nu $ in Figure \ref{fig:twopoint}.
Similarly, the pair correlation function describes the relative probability of
finding a particle at a site, if we know that there is a particle at the other
site. This four-point correlation function is given by the mathematical
expression
\begin{align}
\begin{aligned}
g^{(2)}_{j_1,s_1,j_2,s_2;j_3,s_3,j_4,s_4} = \frac{1}{Z} \tr\rl{ a_{j_1,s_1}^\dagger a_{j_2,s_2}^\dagger a_{j_3,s_3} a_{j_4,s_4} e^{-\beta H} }
\end{aligned}
\end{align}
and can also be evaluated exactly. The full list of four-point correlation 
functions is given in Appendix \ref{correlationf}. In Figure 
\ref{fig:fourpoint} we show a few 
cases of how the four-point correlation functions depend upon the statistics 
parameter $\nu $.
\begin{figure}[h]
	\centering
	\includegraphics[scale=0.4]{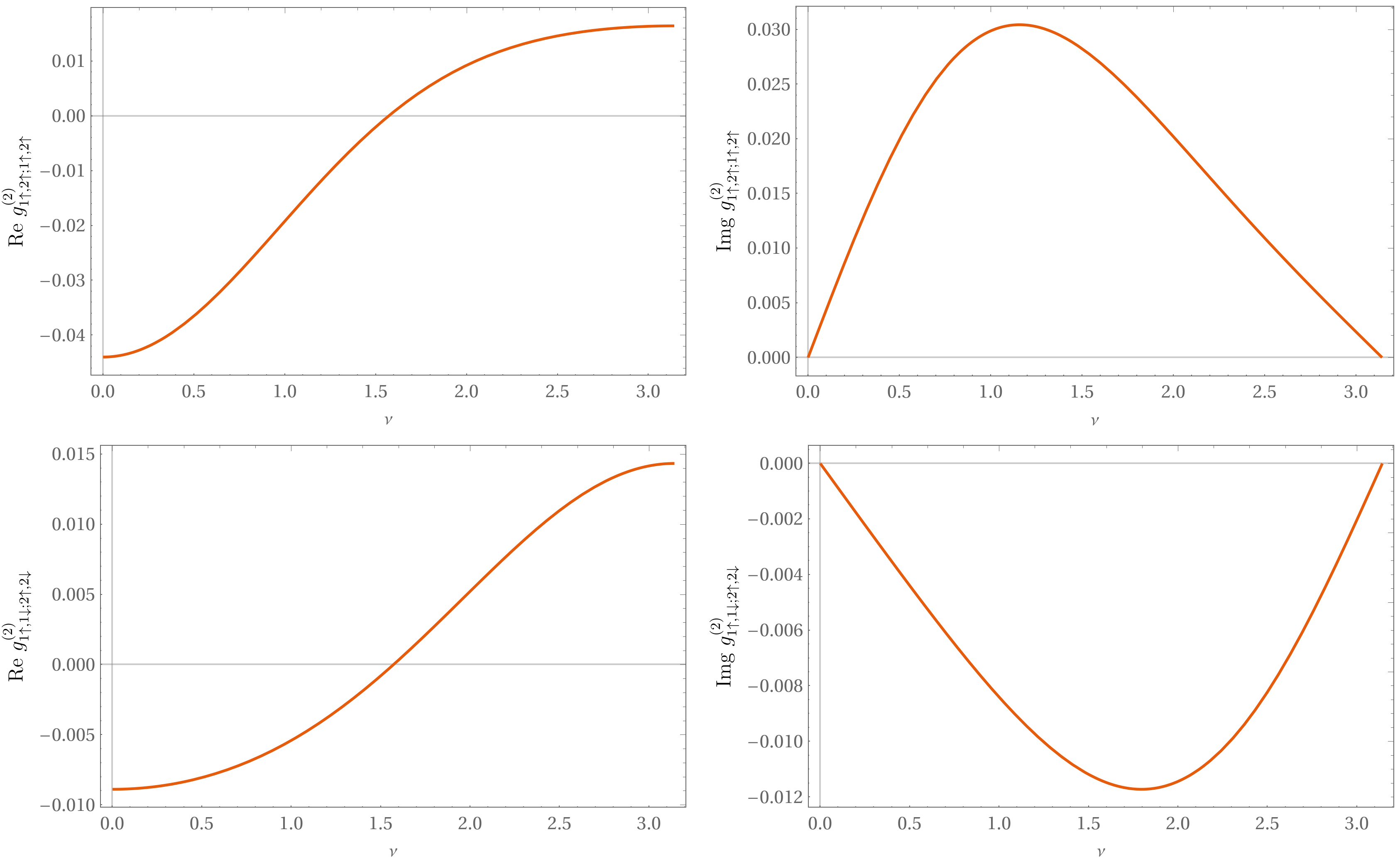}
	\caption{Plots showing dependence of four-point correlation functions on the statistical parameter($\kappa=1,U=4,V=1,J=0.25,\mu=0.5,\beta=1 $). The real part and the imaginary part are separately plotted.}
	\label{fig:fourpoint}
\end{figure}

\section{Quasi-momentum Distribution}\label{quamd}
\begin{figure}[H]
	\centering
	\includegraphics[scale=0.7]{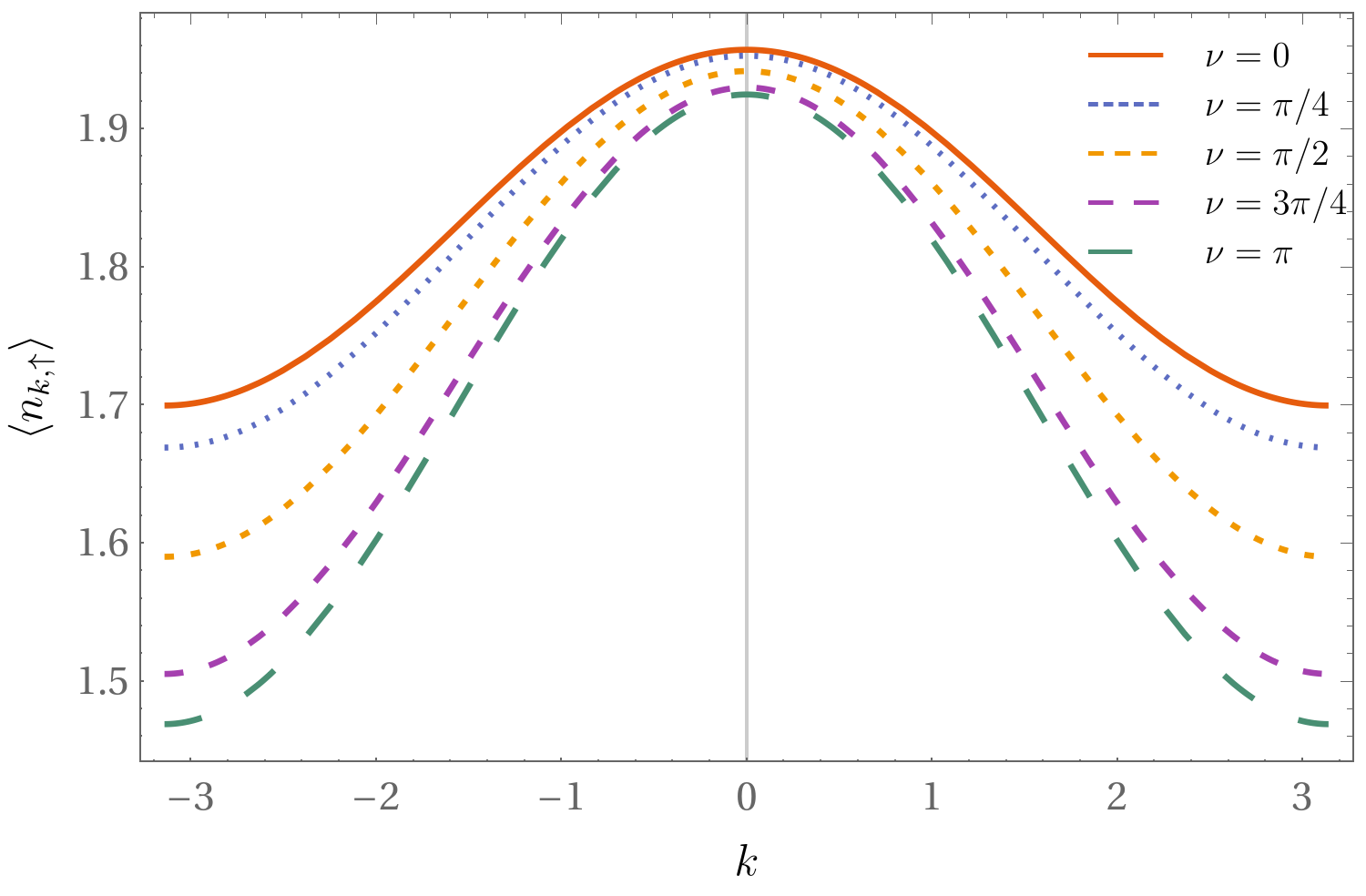}
	\caption{Quasi-momentum distribution is plotted against the momentum for different values of the statistical parameter ($\kappa=1,U=4,V=1,J=0.25,\mu=10,\beta=1 $).}
	\label{fig:momentum}
\end{figure}

The distribution of anyons in the quasi-momentum space is given by the Fourier
transform of the correlation function
\begin{align}
\begin{aligned}
\braket{n_{k,s}} = \sum_{j,j'} e^{i k \rl{j-j'}} g^{(1)}_{j,s;j',s}
\end{aligned}
\end{align}
where $ \braket{n_{ks}} $ denotes the quasi-momentum distribution of spin-$ s $
particles. Since the exact expressions for the correlation functions are known,
it is straightforward to calculate the quasi-momentum distribution function.
Note that since $g^{(1)}_{j\uparrow;j'\uparrow}= g^{(1)}_{j\downarrow;j'
\downarrow}$, the quasi-momentum distribution is same for spin-up and
spin-down particles. In Figure \ref{fig:momentum}, we plot the quasi-momentum distribution for
various values of the statistics parameter.

\section{Summary and Outlook}\label{sumao}
The entanglement of indistinguishable particles poses subtle challenges not
present in the case of distinguishable particles. Ideas based on information
theory come in handy to define a consistent scheme for defining standard
tools like partial trace and reduced density matrix, and correspondingly
measures like von Neumann entropy, even for indistinguishable particles.

When phrased in the language of second quantization, these ideas can
further be extended beyond bosons and fermions -- the context in which
they were developed -- to anyons, particles whose statistics is defined
by a continuous parameter which allows one to interpolate between the
bosonic and fermionic limits.

In this paper,built on an earlier paper 
\cite{maniQuantumEntanglementOnedimensional2020} where anyonic statistics 
were realised as a consequence of the topology of the underlying space, the foregoing points were explored and reinforced in detail
in the concrete case of the one-dimensional two-site anyonic Hubbard model. As a spin-off, some miscellaneous results concerning the one-particle
density matrix, the pair correlation functions of anyons, their
quasi-momentum distribution functions are derived with special emphasis on 
their dependence on the statistics parameter.

Whether the entanglement of anyons discussed in this paper is \textit{protected} \cite{piccoliniEntanglementRobustnessSpatial2021} \cite{nosratiRobustEntanglementPreparation2020} \cite{nosratiDynamicsSpatiallyIndistinguishable2020} is an \textit{open} question, and requires coupling the anyonic Hubbard model to an environment. We hope to discuss this issue elsewhere.

Although in this paper anyonic statistics are introduced  by hand into the 
definition of the Hubbard model, it is possible to relate the anyonic Hubbard 
model with the usual bosonic or fermionic Hubbard models through a 
Jordan-Wigner transformation \cite{keilmannStatisticallyInducedPhase2011}. 
The continuum limit of this model, in turn, is related to the 
Tomonaga-Luttinger liquid with a charge-vortex composite playing the role of 
an anyon \cite{haExactDynamicalCorrelation1994}. It is also well-known that the Tomonaga-Luttinger liquid 
is related to the Calogero-Sutherland model which in itself is an effective 
theory of the fractional quantum Hall system in the lowest Landau level 
\cite{ouvryMappingCalogeroModel2018}. Clearly, there is a commonality in the anyonic 
properties of all these models. It is also of interest to study non-abelian generalisations of 
these results. A related issue worth exploring concerns entanglement in 
systems of indistinguishable particles satisfying exclusion statistics 
\cite{haldaneFractionalStatisticsArbitrary1991}. We hope to examine some of these issues in the near future.
\appendix
\section{Basis Transformation}\label{basistrans}
Consider a projective measurement on to the one-particle subspace 
$\mathcal{B}^{(1)}_M $ where the one-particle basis states are localized 
in $M$. Then, by definition, the reduced density matrix is
\begin{align}
\rho^{(N-1)}_M \rl{\Phi} = \frac{1}{\mathcal{N}_M}\sum_{k\in\mathcal{B}^{(1)}_M} 
a_{k} \ket{\Phi}\bra{\Phi} a^\dagger_{k}
\end{align}
where $\mathcal{N}_M = \braket{\Phi|\sum_{k\in\mathcal{B}^{(1)}_M} 
a^\dagger_{k}a_{k}|\Phi}$. Consider a unitary transformation $ U $ in 
the one-particle subspace $\mathcal{B}^{(1)}_M $ and the basis transformation
\begin{align}
a^\dagger_{k} = \sum_{\tilde{k}} U_{k \tilde{k}} b^\dagger_{\tilde{k}}, 
\quad  \ket{k} \in \mathcal{B}^{(1)}_M 
\end{align}
Corresponding to the one-particle basis states $\ket{k} \notin 
\mathcal{B}^{(1)}_M $, we may choose $a^\dagger_{k} = b^\dagger_{\tilde{k}}$. 
However, these states do not appear in the expression for partial trace onto 
$\mathcal{B}^{(1)}_M$.

We have
\begin{align}
\begin{aligned}
\sum_{k \in \mathcal{B}^{(1)}_M } a_{k} \ket{\Phi}\bra{\Phi} a^\dagger_{k} & = 
\sum_{k,i,j \in \mathcal{B}^{(1)}_M} U_{k,i}^* U_{k,j} b_i\ket{\Phi}\bra{\Phi} 
b_j^\dagger \\
&=\sum_{j\in \mathcal{B}^{(1)}_M} b_j \ket{\Phi} \bra{\Phi} b_j^\dagger 
\end{aligned}
\end{align}
and 
\begin{align}
\begin{aligned}
\mathcal{N}_M = \braket{\Phi|\sum_{k\in\mathcal{B}^{(1)}_M} a^\dagger_{k}a_{k}|
\Phi} =\braket{\Phi|\sum_{k,i,j\in\mathcal{B}^{(1)}_M} U_{k,i}^* U_{k,j}
b_i^\dagger b_j|\Phi }=\braket{\Phi|\sum_{j\in\mathcal{B}^{(1)}_M} b_j^\dagger 
b_j|\Phi }
\end{aligned}
\end{align}
Therefore, the reduced density matrix is given by 
\begin{align}
\rho^{(N-1)}_M\rl{\Phi} = \frac{1}{\mathcal{N}_M}\sum_{k\in\mathcal{B}^{(1)}_M} 
b_{k}\ket{\Phi}\bra{\Phi} b^\dagger_{k}
\end{align}
where $\mathcal{N}_M = \braket{\Phi|\sum_{k\in\mathcal{B}^{(1)}_M}b^\dagger_{k}
b_{k}|\Phi}$. This shows the definition of the reduced density matrix 
is independent of the choice of the basis of the one-particle subspace onto 
which the projective measurement is made.
\section{Eigenvalues and Eigenvectors of the Hamiltonian}\label{eigen}
The eigenvalues of the Hamiltonian are listed below.
\begin{align}\label{eigenvalues}
\begin{aligned}
\varepsilon_{0,1} &=0, \\ \varepsilon_{1,1} &=\varepsilon_{1,3} = -\kappa -\mu\\
\varepsilon_{1,2} &=\varepsilon_{1,4} = \kappa -\mu \\
\varepsilon_{2,1}&=\varepsilon_{2,3}=\varepsilon_{2,6}=2 (-J\cos(\nu)-\mu +V)\\
\varepsilon_{2,2}& = U-2\mu\\
\varepsilon_{2,4}&= \frac{1}{2} \left(-4 \mu + \Upsilon+v_1\right)\\
\varepsilon_{2,5}&= \frac{1}{2} \left(-4 \mu -\Upsilon+v_1\right)\\
\varepsilon_{3,1} &=\varepsilon_{3,3}=-\kappa -2 J \cos (\nu )-3 \mu +U+4 V\\
\varepsilon_{3,2} &=\varepsilon_{3,4}=\kappa -2 J \cos (\nu )-3 \mu +U+4 V\\
\varepsilon_{4,1}& =2 (-2 J \cos (\nu )-2 \mu +U+4 V)
\end{aligned}
\end{align}
where $ v_1= 2 J \cos (\nu )+U+2 V$, $v_2= -2 J \cos
(\nu )+U-2 V$ and $ \Upsilon = \sqrt{\left(16 \kappa
^2+v_2^2\right)} $. $\varepsilon_{j,k}$ denotes the $k^{th}$ eigenvalue of 
the $j-$particle Hamiltonian. The corresponding eigenvectors are
\begin{align}
\begin{aligned}
\ket{\phi_{0,1}} &= \ket{0}\\ 
\ket{\phi_{1,1}}&=\ket{1\uparrow}+\ket{2\uparrow}\\ 
\ket{\phi_{1,2}}&=-\ket{1\uparrow}+\ket{2\uparrow}\\  
\ket{\phi_{1,3}}&=\ket{1\downarrow}+\ket{2\downarrow}\\  
\ket{\phi_{1,4}}&=-\ket{1\downarrow}+\ket{2\downarrow}\\  
\ket{\phi_{2,1}}&=\ket{1\uparrow,2\uparrow}\\  
\ket{\phi_{2,2}}&= -e^{-i \nu }\ket{1\uparrow,1\downarrow}+
\ket{2\uparrow,2\downarrow}\\  
\ket{\phi_{2,3}}&=\ket{1\uparrow,2\downarrow}+\ket{1\downarrow,2\uparrow}\\  
\ket{\phi_{2,4}}&=\frac{4 \kappa e^{-i \nu }}{\Upsilon- v_2}
\ket{1\uparrow,1\downarrow}+\frac{4 \kappa  }{ \Upsilon-v_2}
\ket{2\uparrow,2\downarrow}-\ket{1\uparrow,2\downarrow}+
\ket{1\downarrow,2\uparrow} \\  
\ket{\phi_{2,5}}&=-\frac{4 \kappa e^{-i \nu }}{\Upsilon+ v_2}
\ket{1\uparrow,1\downarrow}-\frac{4 \kappa  }{\Upsilon+ v_2}
\ket{2\uparrow,2\downarrow}-\ket{1\uparrow,2\downarrow}+
\ket{1\downarrow,2\uparrow}\\  
\ket{\phi_{2,6}}&= \ket{1\downarrow,2\downarrow}\\  
\ket{\phi_{3,1}}&= -e^{-i \nu }\ket{1\uparrow,1\downarrow,2\uparrow}+
\ket{1\uparrow,2\uparrow,2\downarrow}\\  
\ket{\phi_{3,2}}&=e^{-i \nu }\ket{1\uparrow,1\downarrow,2\uparrow}+
\ket{1\uparrow,2\uparrow,2\downarrow}\\  
\ket{\phi_{3,3}}&=-e^{-i \nu \
}\ket{1\uparrow,1\downarrow,2\downarrow}+
\ket{1\downarrow,2\uparrow,2\downarrow}\\  
\ket{\phi_{3,4}}&=e^{-i \nu }\ket{1\uparrow,1\downarrow,2\downarrow}+
\ket{1\downarrow,2\uparrow,2\downarrow}\\  
\ket{\phi_{4,1}}&=\ket{1\uparrow,1\downarrow,2\uparrow,2\downarrow}\\  
\end{aligned}
\end{align}

\section{Reduced Density Matrices}\label{reducedmatrices}

Let $ \mathcal{B}_1 = \lbrace \ket{1\uparrow}, \ket{1\downarrow} \rbrace $
and $ \mathcal{B}_2 = \lbrace \ket{2\uparrow}, \ket{2\downarrow} \rbrace $ be
the single-particle local bases at sites 1 and 2 respectively, and let the 
non-local basis be $\mathcal{B}_{12} = \lbrace \ket{1\uparrow}, 
\ket{1\downarrow}, \ket{2\uparrow}, \ket{2\downarrow} \rbrace$. These will
be used to perform the partial traces. 

The reduced density matrices after performing a partial trace 
onto the one-particle subspace $\mathcal{B}_1 = \lbrace \ket{1\uparrow}, 
\ket{1\downarrow} \rbrace $  are
\begin{align}
\begin{aligned}
\rho^{(1)}_{1}\rl{\phi_{2,1}}&=\ket{2\uparrow}\bra{2\uparrow}\\  
\rho^{(1)}_{1}\rl{\phi_{2,2}}&=\frac{1}{2} \rl{\ket{1\uparrow}\bra{1\uparrow}+
\ket{1\downarrow}\bra{1\downarrow}}\\  \rho^{(1)}_{1}
\rl{\phi_{2,3}}&=\frac{1}{2} \rl{\ket{2\uparrow}\bra{2\uparrow}+
\ket{2\downarrow}\bra{2\downarrow}}\\  \rho^{(1)}_{1}\rl{\phi_{2,4}}&=
\frac{\Upsilon +v_2}{4 \Upsilon }\rl{\ket{1\uparrow}\bra{1\uparrow} +
\ket{1\downarrow}\bra{1\downarrow} }-\frac{\kappa  
e^{-i \nu }}{\Upsilon }\rl{\ket{1\uparrow}\bra{2\uparrow}+\ket{1\downarrow}
\bra{2\downarrow}}\\&\quad -\frac{\kappa 
e^{i \nu }}{\Upsilon } \rl{\ket{2\uparrow}\bra{1\uparrow} +\ket{2\downarrow}
\bra{1\downarrow}}+\frac{\Upsilon -v_2}{4 \Upsilon }
\rl{\ket{2\uparrow}\bra{2\uparrow}+\ket{2\downarrow}\bra{2\downarrow} }\\  
\rho^{(1)}_{1}\rl{\phi_{2,5}}&=
\frac{\Upsilon -v_2}{4 \Upsilon } \rl{\ket{1\uparrow}\bra{1\uparrow} +
\ket{1\downarrow}\bra{1\downarrow} } +\frac{\kappa  
e^{-i \nu }}{\Upsilon }\rl{\ket{1\uparrow}\bra{2\uparrow}+
\ket{1\downarrow}\bra{2\downarrow}}\\&\quad+\frac{\kappa  e^{i \nu }}
{\Upsilon } \rl{\ket{2\uparrow}\bra{1\uparrow}+
\ket{2\downarrow}\bra{1\downarrow}}+\frac{\Upsilon +v_2}{4 \Upsilon }
\rl{\ket{2\uparrow}\bra{2\uparrow}+
\ket{2\downarrow}\bra{2\downarrow}}\\  \rho^{(1)}_{1}\rl{\phi_{2,6}}&=
\ket{2\downarrow}\bra{2\downarrow}\\  
\end{aligned}
\end{align}
where $ v_1= 2 J \cos (\nu )+U+2 V$, $v_2= -2 J \cos
(\nu )+U-2 V$ and $ \Upsilon = \sqrt{\left(16 \kappa
	^2+v_2^2\right)} $. 
The reduced density matrices after performing a partial trace 
onto the one-particle subspace $\mathcal{B}_2 = \lbrace \ket{2\uparrow}, 
\ket{2\downarrow} \rbrace $  are
\begin{align}
\begin{aligned}
\rho^{(1)}_{2}\rl{\phi_{2,1}}&=\ket{1\uparrow}\bra{1\uparrow}\\  
\rho^{(1)}_{2}\rl{\phi_{2,2}}&=\frac{1}{2}\rl{\ket{2\uparrow}\bra{2\uparrow}+
\ket{2\downarrow}\bra{2\downarrow}}\\  
\rho^{(1)}_{2}\rl{\phi_{2,3}}&=\frac{1}{2}\rl{\ket{1\uparrow}\bra{1\uparrow}+
\ket{1\downarrow}\bra{1\downarrow}}\\  \rho^{(1)}_{2}\rl{\phi_{2,4}}
&=\frac{\Upsilon -v_2}{4 \Upsilon }\rl{\ket{1\uparrow}\bra{1\uparrow}+\ket{1
\downarrow}\bra{1\downarrow}}-\frac{\kappa
e^{-i \nu }}{\Upsilon }\rl{\ket{1\uparrow}\bra{2\uparrow}+\ket{1\downarrow}
\bra{2\downarrow}}\\& \quad 
-\frac{\kappa  e^{i \nu }}{\Upsilon }
\rl{\ket{2\uparrow}\bra{1\uparrow}+\ket{2\downarrow}\bra{1\downarrow}}+
\frac{\Upsilon +v_2}{4 \Upsilon }
\rl{\ket{2\uparrow}\bra{2\uparrow}+\ket{2\downarrow}\bra{2\downarrow}}\\  
\rho^{(1)}_{2}\rl{\phi_{2,5}}
&=\frac{\Upsilon +v_2}{4 \Upsilon }\rl{\ket{1\uparrow}\bra{1\uparrow}+
\ket{1\downarrow}\bra{1\downarrow}}+\frac{\kappa
e^{-i \nu }}{\Upsilon }\rl{\ket{1\uparrow}\bra{2\uparrow}+\ket{1\downarrow}
\bra{2\downarrow}}\\& \quad 
+\frac{\kappa  e^{i \nu }}{\Upsilon }
\rl{\ket{2\uparrow}\bra{1\uparrow}+\ket{2\downarrow}\bra{1\downarrow}}+
\frac{\Upsilon -v_2}{4 \Upsilon }
\rl{\ket{2\uparrow}\bra{2\uparrow}+\ket{2\downarrow}\bra{2\downarrow}}\\
 \rho^{(1)}_{2}\rl{\phi_{2,6}}&=
\ket{1\downarrow}\bra{1\downarrow}\\  
\end{aligned}
\end{align}
Similarly, the reduced density matrices after performing a 
partial trace onto the one-particle subspace $ \mathcal{B}_{12} = \lbrace 
\ket{1\uparrow}, \ket{1\downarrow}, \ket{2\uparrow}, \ket{2\downarrow}\rbrace$ 
are 
\begin{align}
\begin{aligned}
\rho^{(1)}_{12}\rl{\phi_{2,1}}&=\frac{1}{2}\rl{\ket{1\uparrow}\bra{1\uparrow}+
\ket{2\uparrow}\bra{2\uparrow}} \\  \rho^{(1)}_{12}\rl{\phi_{2,2}}&=
\frac{1}{4}\rl{\ket{1\uparrow}\bra{1\uparrow}+\ket{2\uparrow}
\bra{2\uparrow}+\ket{1\downarrow}\bra{1\downarrow}+
\ket{2\downarrow}\bra{2\downarrow}} \\  
\rho^{(1)}_{12}\rl{\phi_{2,3}}&=\frac{1}{4}\rl{\ket{1\uparrow}\bra{1\uparrow}+
\ket{2\uparrow}\bra{2\uparrow}+\ket{1\downarrow}
\bra{1\downarrow}+\ket{2\downarrow}\bra{2\downarrow}} \\  
\rho^{(1)}_{12}\rl{\phi_{2,4}}
&=\frac{1}{4}\rl{\ket{1\uparrow}\bra{1\uparrow}+\ket{1\downarrow}
\bra{1\downarrow}}-\frac{\kappa
e^{-i \nu }}{\Upsilon }\rl{\ket{1\uparrow}\bra{2\uparrow}+\ket{1\downarrow}
\bra{2\downarrow}} \\& \quad 
-\frac{\kappa  e^{i \nu }}{\Upsilon }
\rl{\ket{2\uparrow}\bra{1\uparrow}+\ket{2\downarrow}\bra{1\downarrow}}+
\frac{1}{4  }
\rl{\ket{2\uparrow}\bra{2\uparrow}+\ket{2\downarrow}\bra{2\downarrow}}\\  
\rho^{(1)}_{12}\rl{\phi_{2,5}}
&=\frac{1}{4}\rl{\ket{1\uparrow}\bra{1\uparrow}+\ket{1\downarrow}
\bra{1\downarrow}}+\frac{\kappa
e^{-i \nu }}{\Upsilon }\rl{\ket{1\uparrow}\bra{2\uparrow}+\ket{1\downarrow}
\bra{2\downarrow}} \\ & \quad 
+\frac{\kappa  e^{i \nu }}{\Upsilon }
\rl{\ket{2\uparrow}\bra{1\uparrow}+\ket{2\downarrow}\bra{1\downarrow}}+
\frac{1}{4}
\rl{\ket{2\uparrow}\bra{2\uparrow}+\ket{2\downarrow}\bra{2\downarrow}}  \\
\rho^{(1)}_{12}\rl{\phi_{2,6}}&=\frac{1}{2}\rl{\ket{1\downarrow}
\bra{1\downarrow}+\ket{2\downarrow}\bra{2\downarrow}}   
\end{aligned}
\end{align}

\section{Time Evolution in the Presence of an External Field}\label{coeff}

From the block structure of the Hamiltonian, it is clear that in the presence 
of an external field the state $ \ket{\Phi_i^{(2)}}  = \ket{1\uparrow,1
\downarrow}$ is evolved by a time evolution operator $ \tilde{W}(t) $ which is the 
solution of the equation
\begin{align}
\begin{aligned}
i \dfrac{d \tilde{W}(t)}{dt} = H_{2,0}(t) \tilde{W}(t)
\end{aligned}
\end{align}
where 
\begin{align}
\begin{aligned}
H_{2,0}(t)  &=\left(
\begin{array}{cccc}
U-2\mu &0& -\kappa e^{-i (\nu- \Omega t)} & \kappa  e^{-i (\nu-\Omega t)} \\
0 & U-2 \mu  &  - \kappa e^{-i \Omega t
}& \kappa  e^{-i \Omega t } \\
-  \kappa e^{i (\nu -\Omega t )} &
- \kappa e^{i \Omega t }& 2
(V-\mu ) & -2 J \cos (\nu ) \\
\kappa  e^{i (\nu -\Omega t )} & \kappa  e^{i 
	\Omega t } & -2 J \cos (\nu ) & 2 (V-\mu ) \\
\end{array}
\right)
\end{aligned}
\end{align}
Here the matrix representation of the Hamiltonian is given in the basis 
\[ \lbrace \ket{1\uparrow,1\downarrow}, \ket{2\uparrow,2\downarrow}, 
\ket{1\uparrow,2\downarrow},\ket{1\downarrow,2\uparrow} \rbrace .\]

The time evolution operator $ \tilde{W}(t)$ can be found as discussed in section 
\ref{extfield} and is given by $ \tilde{W}(t) = W_1 e^{-i \tilde{H}_{2,0} t} $ 
where $W_1 = \text{diag}\lbrace e^{- i\Omega t}, e^{i\Omega t},1,1\rbrace$ 
and 
\begin{align}
\begin{aligned}
\tilde{H}_{2,0} =\left(
\begin{array}{cccc}
U-2 \mu +\Omega  & 0 &  -\kappa e^{-i \nu
} & \kappa  e^{-i \nu } \\
0 & U-2 \mu -\Omega  & -\kappa  & \kappa  \\
-\kappa e^{i \nu } & -\kappa  & 2
(V-\mu ) & -2 J \cos (\nu ) \\
\kappa  e^{i \nu } & \kappa  & -2 J \cos (\nu )
& 2 (V-\mu ) \\
\end{array}
\right)
\end{aligned}
\end{align}
Even though we can obtain  $\tilde{W}(t)$ exactly, we will not discuss it here as the 
expressions are long and cumbersome. However we will give the following 
argument to establish certain relations among the matrix elements of $\tilde{W}(t)$.

In the presence of the external field, the state $\ket{\Phi_i^{(2)}}$ evolves 
into a state of the form
\begin{align}
\begin{aligned}
\ket{\Phi_o^{(2)}} = \tilde{W}(t) \ket{\Phi_i^{(2)}} = \tilde{W}_{11}(t)\ket{1\uparrow,1\downarrow}+\tilde{W}_{12}(t) \ket{2\uparrow,2\downarrow}+\tilde{W}_{13}(t) \ket{1\uparrow,2\downarrow}+\tilde{W}_{14}(t)\ket{1\downarrow,2\uparrow}
\end{aligned}
\end{align}
where $\tilde{W}_{m,n}(t)$ are matrix elements of $\tilde{W}(t)$. Since $\tilde{W}(t)$ is a unitary 
operator, it is clear that $|\tilde{W}_{11}(t)|^2+|\tilde{W}_{12}(t)|^2+|\tilde{W}_{13}(t)|^2+
|\tilde{W}_{14}(t)|^2=1$.

Note that the Hamiltonian $ \tilde{H}_{2,0} $ is invariant under the unitary 
similarity transformation \[ P\tilde{H}_{2,0} P^\dagger  = \tilde{H}_{2,0}\] 
where
\begin{align}
\begin{aligned}
P = \left(
\begin{array}{cccc}
1 & 0 & 0 & 0 \\
0 & 1 & 0 & 0 \\
0 & 0 & 0 & -1 \\
0 & 0 & -1 & 0 \\
\end{array}
\right)
\end{aligned}
\end{align}
This, in turn, also means that $ \tilde{W}(t) $ is invariant under the unitary 
similarity transformation by the matrix $ P $. Therefore we can conclude that 
$ \tilde{W}_{14}(t) = -\tilde{W}_{13}(t) $ resulting in the output state 
\begin{align}
\begin{aligned}
\ket{\Phi_o^{(2)}} = \tilde{W}_{11}(t)\ket{1\uparrow,1\downarrow}+\tilde{W}_{12}(t) 
\ket{2\uparrow,2\downarrow}+\tilde{W}_{13}(t) \rl{\ket{1\uparrow,2\downarrow}-
\ket{1\downarrow,2\uparrow}}
\end{aligned}
\end{align}
Identifying $c_1(t) =\tilde{W}_{11}(t), c_2(t) =\sqrt{2}\tilde{W}_{13}(t)$ and $c_3(t) = 
\tilde{W}_{12}(t)$ we obtain the result in equation \ref{outstate}.

\section{Partition Function}\label{partitionf}

The partition function is easily calculated and is given below. 
\begin{align}
\begin{aligned}
Z &=1+2 e^{\beta  (\kappa +\mu )}+2 e^{-\beta 
	(\kappa -\mu )}+2 e^{-\beta  (-\kappa +U+4 V -2 J
	\cos (\nu )-3 \mu)}+2 e^{-\beta 
	(\kappa+U+4 V -2 J \cos (\nu )-3 \mu )}\\&+e^{-2
	\beta  (U+4 V-2 J \cos (\nu )-2 \mu )}+3
e^{-2 \beta  (V-J \cos (\nu )-\mu
	)}+e^{-\beta  (U-2 \mu )}+e^{-\frac{1}{2}
	\beta  \left(-4 \mu -\sqrt{16 \kappa
		^2+v_2^2}+v_1\right)}\\&+e^{-\frac{1}{2} \beta 
	\left(-4 \mu +\sqrt{16 \kappa
		^2+v_2^2}+v_1\right)}
\end{aligned}
\end{align}
where $ v_1= 2 J \cos (\nu )+U+2 V$, $v_2= -2 J \cos
(\nu )+U-2 V$ and $ \Upsilon = \sqrt{\left(16 \kappa
^2+v_2^2\right)} $. Note that the partition function reduces to 
the fermionic case in the limit $ \nu=0 $ and to the (pseudo-)bosonic 
case in the limit $ \nu=\pi $. 

\section{Correlation Functions}\label{correlationf}
The correlation functions can be calculated exactly since the exact solution 
of the two-site Anyonic Hubbard Model are known. The two-point correlation 
functions are 
\begin{align}
\begin{aligned}
g^{(1)}_{i,s;i,s} =& \frac{1}{2Z} \bigg(e^{-\beta  \varepsilon
	_{1,1}}+e^{-\beta  \varepsilon _{1,2}}+3
e^{-\beta  \varepsilon _{2,1}}+e^{-\beta 
	\varepsilon _{2,2}}\\&+e^{-\beta  \varepsilon
	_{2,4}}+e^{-\beta  \varepsilon _{2,5}}+3
e^{-\beta  \varepsilon _{3,1}}+3 e^{-\beta 
	\varepsilon _{3,2}}+2 e^{-\beta  \varepsilon
	_{4,1}}\bigg) \\
g^{(1)}_{i,s;j,s}&= \frac{1}{2Z} \left(-\frac{4 \kappa  e^{-\beta 
		\varepsilon _{2,4}}}{\Upsilon }+\frac{4 \kappa
	e^{-\beta  \varepsilon _{2,5}}}{\Upsilon
}+e^{-\beta  \varepsilon _{1,1}}-e^{-\beta 
	\varepsilon _{1,2}}+e^{-\beta  \varepsilon
	_{3,1}}-e^{-\beta  \varepsilon _{3,2}}\right), \quad i \neq j \\
g^{(1)}_{i,s;j,s'} & = 0, \quad s \neq s' 
\end{aligned}
\end{align}
where $ \varepsilon_{m,n} $ is the $ n^{\text{th}} $ eigenvalue of $m-$particle
Hamiltonian, $ i,j=1,2 $ and $ s,s'=\uparrow,\downarrow $. 

Similarly, the four-point correlation functions can be exactly calculated. The distinct cases of the four-point correlation functions are
\begin{align}
\begin{aligned}
g^{(2)}_{1\uparrow,2\uparrow;1\uparrow,2\uparrow} =&e^{-i \nu } \left(-e^{-\beta\varepsilon
	_{2,1}}-e^{-\beta  \varepsilon
	_{3,1}}-e^{-\beta  \varepsilon
	_{3,2}}-e^{-\beta  \varepsilon _{4,1}}\right)
 \\
g^{(2)}_{1\uparrow,1\downarrow;1\uparrow,1\downarrow} =& \frac{1}{4} \bigg(-2 e^{-\beta\varepsilon
	_{2,2}}-e^{-\beta  \varepsilon
	_{2,4}}-e^{-\beta  \varepsilon _{2,5}}-4
e^{-\beta  \varepsilon _{3,1}}-4 e^{-\beta 
	\varepsilon _{3,2}}-4 e^{-\beta  \varepsilon
	_{4,1}}  \\&+\frac{v_2 \left(e^{-\beta  \varepsilon
		_{2,5}}-e^{-\beta  \varepsilon
		_{2,4}}\right)}{\Upsilon }\bigg)
 \\
g^{(2)}_{1\uparrow,1\downarrow;2\uparrow,2\downarrow} =& \frac{1}{4}e^{i\nu} 
\left(2 e^{-\beta 
	\varepsilon _{2,2}}-e^{-\beta  \varepsilon
	_{2,4}}-e^{-\beta  \varepsilon
	_{2,5}}+\frac{v_2 \left(e^{-\beta  \varepsilon
		_{2,5}}-e^{-\beta  \varepsilon
		_{2,4}}\right)}{\Upsilon }\right)
 \\
g^{(2)}_{1\uparrow,1\downarrow;1\uparrow,2\downarrow}= &  \frac{2 \kappa  e^{-\beta  \varepsilon _{2,4}}-2
	\kappa  e^{-\beta  \varepsilon
		_{2,5}}-\Upsilon  e^{-\beta  \varepsilon
		_{3,1}}+\Upsilon  e^{-\beta  \varepsilon
		_{3,2}}}{2 \Upsilon } \\
g^{(2)}_{1\uparrow,2\downarrow;1\uparrow,2\downarrow}=& \frac{1}{4} e^{-i \nu } \bigg(-2 e^{-\beta 
	\varepsilon _{2,1}}-4 e^{-\beta  \varepsilon
	_{3,1}}-4 e^{-\beta  \varepsilon _{3,2}}-4
e^{-\beta  \varepsilon
	_{4,1}} \\&+\frac{\left(v_2-\Upsilon \right)
	e^{-\beta  \varepsilon _{2,4}}}{\Upsilon
}-\frac{\left(\Upsilon +v_2\right) e^{-\beta 
		\varepsilon _{2,5}}}{\Upsilon }\bigg) \\
g^{(2)}_{1\uparrow,2\downarrow;1\downarrow,2\uparrow}=& \frac{e^{-i\nu }}{4 \Upsilon } 
\left(-2 \Upsilon  e^{-\beta 
	\varepsilon _{2,1}}+\left(\Upsilon -v_2\right)
e^{-\beta  \varepsilon _{2,4}}+\left(\Upsilon
+v_2\right) e^{-\beta  \varepsilon
	_{2,5}}\right) \\
g^{(2)}_{j_1,s_1,j_2,s_2;j_3,s_3,j_4,s_4} =&0 \text{  if  } s_1+s_2 \neq s_3+s_4 
\end{aligned}
\end{align}
We can obtain the other four-point correlation functions using the following relations
\begin{align}
\begin{aligned}
g^{(2)}_{1\uparrow,2\uparrow;1\uparrow,2\uparrow} &=g^{(2)}_{1\downarrow,2
	\downarrow;1\downarrow,2\downarrow} \\
g^{(2)}_{1\uparrow,1\downarrow;1\uparrow,1\downarrow} &=g^{(2)}_{2\uparrow,2
	\downarrow;2\uparrow,2\downarrow} \\
g^{(2)}_{1\uparrow,1\downarrow;1\uparrow,2\downarrow}&=-g^{(2)}_{1\uparrow,1
	\downarrow;1\downarrow,2\uparrow}=e^{-i \nu}g^{(2)}_{2\uparrow,2\downarrow;1
	\uparrow,2\downarrow}=-e^{-i \nu}g^{(2)}_{2\uparrow,2\downarrow;1\downarrow,2
	\uparrow}\\&=e^{-i \nu}g^{(2)}_{1\uparrow,2\downarrow;1\uparrow,1\downarrow}
=g^{(2)}_{1\uparrow,2\downarrow;2\uparrow,2\downarrow}=-e^{-i\nu}g^{(2)}_{1
	\downarrow,2\uparrow;1\uparrow,1\downarrow}=-g^{(2)}_{1\downarrow,2\uparrow;
	2\uparrow,2\downarrow}
\end{aligned}
\end{align}
\bibliographystyle{unsrt}
\bibliography{papthree.bib} 
\end{document}